\documentclass[a4paper,10pt]{article}
\usepackage{jheppub} 
\usepackage{booktabs}
\usepackage{easytable}
\usepackage[utf8]{inputenc}
\usepackage{longtable}
\usepackage[backend=biber,style=numeric-comp,sorting=none]{biblatex}
\usepackage{amsmath}
\usepackage{array}
\usepackage{MnSymbol}
\usepackage{amsthm}
\usepackage{multirow}
\usepackage{multicol}
\usepackage{amssymb}
\usepackage{graphicx}
\usepackage{slashed}
\usepackage{MnSymbol}
\usepackage{mathtools}
\usepackage{cancel}
\allowdisplaybreaks
\usepackage{xcolor}
\usepackage{simpler-wick}
\usepackage{hyperref} 
\hypersetup{backref=true,       
    pagebackref=true,               
    hyperindex=true,                
    colorlinks=true,                
    breaklinks=true,                
    urlcolor= black,                
    linkcolor= blue,                
    bookmarks=true,                 
    bookmarksopen=false,
    filecolor=black,
    citecolor=blue,
    linkbordercolor=blue
}
\AtBeginDocument{\hypersetup{pdfborder={0 0 1}}}
\numberwithin{equation}{section}
\newtheorem{theorem}{Theorem}
\theoremstyle{definition}

\newcommand{\efill}{\;\;\;\;\;\;\;\;\;\;}
\newcommand{\be}{\begin{equation}}

\newcommand{\ee}{\end{equation}}
\newcommand{\bthe}{\begin{theorem}}
\newcommand{\ethe}{\end{theorem}}
\newcommand{\bea}{\begin{eqnarray}}
\newcommand{\eea}{\end{eqnarray}}
\newcommand{\al}{\alpha}
\renewcommand{\d}{\delta}

\newcommand{\G}{\Gamma}
\newcommand{\g}{\gamma}
\renewcommand{\k}{\kappa}
\newcommand{\La}{\Lambda}
\newcommand{\la}{\lambda}
\newcommand{\m}{\mu}
\newcommand{\n}{\nu}

\newcommand{\om}{\omega}
\newcommand{\s}{\sigma}

\newcommand{\hlf}{\frac{1}{2}}

\newcommand{\M}{\mathcal{M}}

\newcommand{\p}{\partial}

\newcommand{\R}{\mathbb{R}}
\newcommand{\rr}{\rightarrow}

\newcommand{\Z}{\mathbb{Z}}

\newcommand{\SL}{\operatorname{SL}}

\newcommand{\Tr}{\operatorname{Tr}}

\newcommand{\lp}{\left(}
\newcommand{\rp}{\right)}
\newcommand{\ls}{\left[}
\newcommand{\rs}{\right]}

\newcommand{\bq}{\bar{q}}

\newcommand{\wt}{\widetilde}

\newcommand{\wtal}{\widetilde{\al}}

\newcommand{\wtpsi}{\widetilde{\psi}}

\newcommand{\ov}[1]{{\overline{#1}}}

\title{\boldmath Six and seven dimensional extrema of  Non-SUSY heterotic theory with non-maximal enhancements}

\author[1]{Stevell Muller}\author[2]{, Daniel Robbins}
\author[2]{and Hassaan Saleem}
\affiliation[1]{Institut für Algebraische Geometrie, Leibniz Universität Hannover, Hannover, Germany}
\affiliation[2]{Department of Physics, University at Albany, 1400 Washington Avenue, Albany, NY, USA}

\abstract{ We propose a construction of non-supersymmetric heterotic theories on $T^n$ for $n=1,2,3,4$ which lead to extrema of the one-loop effective potential.  In contrast to other approaches, we do not impose maximal gauge enhancement, instead relying on Weyl symmetries of the right-moving part of the momentum lattice to ensure that we get an extremum. The classification depends on finding all lattices within a given genus using Kneser's method.  We find several thousand extrema, most of which have non-maximal rank gauge enhancements.  Several hundred of our extrema are tachyon-free, and several dozen are also not knife edges.  There may be further equivalences among these points, but we find at least 2, 11, and 38 distinct tachyon-free non-knife-edge points for $T^2$, $T^3$, and $T^4$, respectively. In addition, we build on our previous work to study the perturbative stability of $AdS_{3}\times S^{3}\times T^{4}$ backgrounds by checking for Breitenlohner-Freedman (BF) bound violations at the tachyon-free non-knife-edge extrema that we found. The lowest eigenvalues of the Hessian at all such extrema that have a positive one loop potential show violations of the BF bound and satisfy the refined de Sitter conjecture.}

\begin{document}
\maketitle
\flushbottom



 

\newpage

\section{Introduction}

There are many good phenomenological motivations for supersymmetry which is broken at a relatively low energy scale.  With these motivations in mind, much of the focus of efforts to build models in superstring theory that might make contact with real-world physics have started with supersymmetric strings and tried to break the spacetime supersymmetry at the compactification or lower scales using ingredients like fluxes, branes, or other mechanisms.  Unfortunately, despite extensive searches for evidence of low-energy spacetime supersymmetry by collider experiments and cosmological observations, such evidence remains elusive.  On the theory side, this suggests that maybe we should turn more of our focus to solutions in which spacetime supersymmetry is absent (or maybe we say it's broken up at the string scale) from the beginning.

Following this approach gives us three known ten-dimensional string theories without the presence of tachyons, where one of these string theories has only closed strings \cite{Alvarez-Gaume:1986ghj} and the other two have open and closed strings \cite{Sagnotti:1995ga, Sugimoto:1999tx}. Note that, although the spacetime supersymmetry is broken, these theories retain some supersymmetry on the worldsheet which helps ensure consistency.  Over the last few decades, real-world model building using these string theories has received much less attention than their supersymmetric cousins, since these theories don't provide the calculational control that comes due to supersymmetry. However, this trend is starting to change, as there has been much interest in the study of these string theories and their relation with other string theories in the last ten years~\cite{Nibbelink_2015, blaszczyk2016calabiyaucompactificationsnonsupersymmetricheterotic, Blaszczyk:2015zta, Basile:2018irz, Antonelli:2019nar, Basile:2020mpt, Baykara:2022cwj, Raucci:2022bjw, Fraiman:2023cpa, Mourad:2024mpg, Dudas:2025ubq,  Raucci_2025, Menet:2025nbf, Robbins:2025wlm, Basile:2026lyc, Altavista_2026, honda2026, Ishige:2026ylw, fraiman2026nonsupersymmetricdualitiesgaugealgebra, nibbelink2016linebundleembeddingsheterotic, defreitas2024nonsupersymmetricheteroticstringschiral, angelantonj2024nonsupersymmetricnontachyonicheteroticvacua, baykara2024newnonsupersymmetrictachyonfreestrings, Basaad:2024lno, Tachikawa_2024, Leone_2026, Basile_2021, Mourad_2021, Basile_2022, basile2024globalanomaliesbordism, larotonda2026anomalyinflowgaugegroup, baykara2026mtheorys1vees1type, Altavista:2026evd}.

In the current work, we consider the non-supersymmetric tachyon-free heterotic theory, known as the $O(16)\times O(16)$ heterotic theory.  Because of the lack of spacetime supersymmetry, the torus (and higher-loop) vacuum amplitude of this theory does not vanish, and leads to a contribution to the effective potential in target space.  In ten dimensions, this is a potential for the dilaton, and it means that flat space with a constant dilaton is no longer a solution.

Since flat space is not a solution, it is natural to consider compactifications of this theory.  The first thing one might try is compactification on $S^1$, and this background was considered by Ginsparg and Vafa~\cite{Ginsparg:1986wr}.  Now in addition to the dilaton, the string spectrum includes seventeen additional scalars (the circle radius and sixteen Wilson lines), and the one-loop potential depends on the vacuum expectation values of these fields.  Although the fields are massless at tree-level, they generically get masses from the one-loop potential, and in this paper we use the term ``pseudo-moduli''.  The potential is surprisingly rich, but one can find extrema of the potential in the pseudo-moduli space~\cite{Ginsparg:1986wr,Fraiman:2023cpa}, for instance when the Wilson lines are set to zero and the radius is at the self-dual radius.  Even at these points, the potential does not stabilize the dilaton, leading to runaway behavior unless we do additional compactification.  One way to stabilize the dilaton is to compactify further on products of $AdS$ spaces and spheres with fluxes turned on.  For instance in ten dimensions we have $AdS_7\times S^3$ \cite{Basile:2018irz}, and $AdS_4\times S^3\times S^3$\cite{Basile:2026lyc} solutions, and in nine dimensions we can construct $AdS_3\times S^3\times S^3\times S^1$ solutions~\cite{Baykara:2022cwj}. Some other $AdS$ solutions with the internal space having a non-trivial fiber structure were explored in \cite{Raucci_2025}. For these sorts of constructions, the goal becomes finding extrema in the pseudo-moduli space and then checking the stability of these solutions.

In this paper, we will focus on finding nonsupersymmetric heterotic theories on $T^n$ for $n\le 4$.  Some examples of this for $n=4$, along with extensive analysis of perturbative stabilities of the corresponding $AdS_3\times S^3\times T^4$ solutions, were described in~\cite{Robbins:2025wlm}. The key task is to find extrema of the one-loop potential in the $(16n+n^2)$-dimensional pseudomoduli space. For compactifications on a torus $T^n$, the set of all such extrema is not known, and there are techniques to generate some classes of them \cite{Font_2020, Fraiman:2023cpa, honda2026}. The extrema in such classes tend to be the points where the gauge group of the compactification acquires a non-abelian factor. We refer to this phenomenon as gauge group enhancement. A significant body of work has been dedicated to finding extrema with maximal enhancement (which means that for a $T^n$ compactification, the non-abelian part of the gauge group has rank $16+n$ and has no Abelian factors) \cite{Fraiman:2023cpa, honda2026}. However, studies to classify (or at least systematically generate examples of) non-maximal enhancements are still fairly open, even though there is a significant phenomenological motivation to do so, as the non-Abelian part of the low-energy gauge group of the Standard Model has rank three.

This work initiates such an effort to systematically generate classes of non-maximal extrema. These examples come from a construction that is similar to the construction used in \cite{honda2026} (which is inspired by \cite{Font_2020} and \cite{Nikulin1980}) and involves two even lattices that we call $\Lambda_L$ and $\Lambda_R$. However, there are significant differences between our approach and the approaches used in the aforementioned references, as we focus on the case where $\Lambda_R$ is an ADE root lattice, and then we find the most general compatible $\La_L$ lattices with which it can pair. We perform the analysis for all possible $ADE$ lattices for $S^1,\;T^2$, and $T^3$ compactifications and three out of six possibilities for $T^4$ compactifications.

This analysis leads to $8619$ examples of extrema, with $7574$ of them having non-maximal enhancements (with one example having a rank of as low as $10$ for $T^4$ compactifications). Out of these extrema, $668$ are free of tachyons, and $113$ of these tachyon-free points aren't knife edges (with $392$ of these tachyon-free extrema having non-maximal enhancements, and $91$ of these non-maximal enhancements not being knife edges). These extrema for $S^1$ are given in Table \ref{tab: rank 1 A1} and for higher tori, they are provided in the ancillary files for this paper. We highlight the extrema with non-maximal enhancements in red. We also calculate the Hessians for the tachyon-free, non-knife-edge extrema, and none of these points turn out to be a local minimum. Building on our previous work \cite{Robbins:2025wlm}, we use the Hessians for $T^4$ extrema to check for the Breitenlohner-Freedman (BF) bound violations \cite{Breitenlohner:1982jf} for the $AdS_3\times S^3\times T^4$ backgrounds studied in \cite{Robbins:2025wlm} in the ``intrinsically quantum'' limit. We also checked that the extrema with a positive potential satisfy the refined de Sitter conjecture \cite{Ooguri:2018wrx, Andriot:2018wzk}.

This paper is organized as follows. In Section \ref{sec 2: lattice construction}, we outline the construction that we use for an even self-dual lattice $\Gamma$ in terms of two even lattices $\Lambda_L$, and $\Lambda_R$, and the argument for getting extrema if we choose $\Lambda_R$ as an $ADE$ lattice and $\d_R=0$. In Section \ref{sec 3: Finding extrema}, we mention the algorithms that we use to get $\Lambda_L$s for a given $\Lambda_R$, and we outline the procedure to get $\d$s for the ten-dimensional case (i.e., $n=0$ case) as an example. We also tabulate the number of extrema (and the number of tachyon-free extrema) with enhancements of different ranks in this section for $T^2, T^3$ and $T^4$ compactifications as well. In Section \ref{sec 4: Hessian}, we outline our procedure to calculate the Hessian, and we check for the existence of BF bound violations of the $AdS_3\times S^3\times T^4$ backgrounds discussed in \cite{Robbins:2025wlm}. We mention our conventions for the $ADE$ lattices and theta functions in Appendices \ref{app: ADE lattices} and \ref{app: theta functions}, provide the lists of the root systems of $\Lambda_L$s with the number of enhancements they get, and the values of the Hessians in Appendix \ref{app: enhancements}.

\section{Lattice construction}\label{sec 2: lattice construction}
\subsection{Toroidal compactifications of supersymmetric heterotic strings}

Let us start by reviewing toroidal compactifications of the supersymmetric heterotic string.  Throughout this paper, we will use the bosonic formulation of the heterotic string.  Then the main ingredient used to specify a toroidal compactification to $10-n$ dimensions is an even self-dual lattice $\Gamma$ of rank $16+2n$ and signature $(16+n,n)$.  In other words, $\Gamma$ is a set of vectors in $\R^{16+n,n}$ that forms an additive subgroup (necessarily isomorphic to $\Z^{16+2n}$) such that $v^2\in 2\Z$ for all $v\in\Gamma$ and with $\Gamma^\vee=\Gamma$, where the dual lattice is defined as
\be
\Gamma^\vee:=\left\{v\in\R^{16+n,n}\mid v\cdot u\in\Z, \forall u\in\Gamma\right\}.
\ee
Let us make some comments about notation.  For a Euclidean vector space or lattice, a bar overtop will denote the same underlying set but with a negative definite inner product.  Thus we have, for instance, $\R^{16+n,n}=\R^{16+n}\oplus\overline{\R^n}$.  When listing the components of a vector in a split-signature vector space, we will divide the positive signature components from the negative signature components by a vertical line, so we would write $v\in\R^{16+n,n}$ as $v=(v^1,\cdots,v^{16+n}\mid v^{17+n},\cdots,v^{16+2n})$.  When writing out a specific vector in components, we will use exponents to indicate repeated components.  Thus we would have, for instance,
\be\label{eq:ComponentVectorExample}
v=\lp\hlf,\hlf,\hlf,\hlf,\hlf,\hlf,\hlf,\hlf,0,0,0,0,0,0,0,0,0\mid\frac{1}{\sqrt{2}}\rp=\lp\hlf^8,0^9\mid\frac{1}{\sqrt{2}}\rp\in\R^{17,1}.
\ee
When we explicitly split our vector into components, the default is to take the individual left and right inner products to be Euclidean, so if we write $v=(v_L|v_R)$ with $v_L=(\hlf^8,0^9)$, $v_R=(\frac{1}{\sqrt{2}})$, then we would say $v_L^2=2$, $v_R^2=\hlf$, and $v^2=v_L^2-v_R^2=\frac{3}{2}$.  Our approach will often be to build lattices up using direct sums of smaller lattices.  To make the distinctions between blocks clearer, we will use a semi-colon instead of a comma to indicate the boundaries between these blocks in a component expression.  So for instance, in the example~\eqref{eq:ComponentVectorExample} above, if this were a vector inside the lattice $\Gamma_{E_8}\oplus\Gamma_{D_9}\oplus\overline{\frac{1}{\sqrt{2}}\Z}$ (whose blocks will be defined below), we would write it as $v=(\hlf^8;0^9|\frac{1}{\sqrt{2}})$.  Finally, for some of our standard lattices (specifically the $A_n$ and the $E_7$ root lattices and their duals) we will find it convenient to embed the lattice in a vector space one dimension higher, restricting to the subspace in which the components sum to zero.  This means that the number of components will not always match the lattice rank, and some expressions must be taken with this context in mind.  As an example, we might have a vector in the rank seventeen lattice $\Gamma_{E_8}\oplus\Gamma_{E_7}^\vee\oplus\Gamma_{A_2}^\vee$ described as $\lp0^8;\frac{1}{4}^6,-\frac{3}{4}^2;\frac{1}{3}^2,-\frac{2}{3}\rp$.

Given our $(16+n,n)$ even self-dual lattice $\Gamma$, we can describe the low-lying spectrum of the corresponding heterotic string background.  For simplicity, we will use a light-cone description, with indices $\m,\n=2,\cdots,9-n$, $i,j=1,\cdots,16+n$, and $a,b=1,\cdots,n$.  There are left-moving bosons with raising operators $\al^\m_{-k}$ and $\al^i_{-k}$ and right-moving bosons with operators $\wtal^\m_{-k}$ and $\wtal^a_{-k}$.  There are also right-moving fermions which can be in either the NS or R sectors and have raising operators $\wtpsi^\m_{-r}$ and $\wtpsi^a_{-r}$, where $r>0$ is either in $\Z$ for the R sector or $\Z+\hlf$ for the NS sector.  States are labeled by a vector $(p_L|p_R)\in\Gamma$ as well as a set of raising operators and a space-time momentum $q$ in $\R^{9-n,1}$. Level matching gives
\be\label{eq:LevelMatching}
m^2=-q^2=\hlf p_L^2+N_L-1=\hlf p_R^2+N_R-a,
\ee
where $m^2$ is the mass of a state corresponding to the spacetime momentum $q$. In addition, $a=\hlf$ is in the NS sector and $a=0$ in the R sector.  There is also a GSO projection which requires an odd number of $\wtpsi$ raising operators in the NS sector and projects onto one chirality in the R sector.  Because of this projection, $N_R-a\ge 0$ in both sectors, and so there are no tachyons.

The massless fields have on the left either $N_L=1$, from either $\al^\m_{-1}$ or $\al^i_{-1}$, and $p_L=0$, or have $N_L=0$ and $p_L^2=2$, and on the right are either R sector with $p_R=0$ and $N_R=0$, or NS sector with $N_R=\hlf$ and $p_R=0$.  The NS sector states $\al^\m_{-1}\wtpsi^\n_{-\hlf}|q,(0|0)\rangle$ give the usual metric, $B$-field, and dilaton in $10-n$ dimensions.  The NS sector states $\al^\m_{-1}\wtpsi^a_{-\hlf}|q,(0|0)\rangle$, $\al^i_{-1}\wtpsi^\m_{-\hlf}|q,(0|0)\rangle$, and $\wtpsi^\m_{-\hlf}|q,(p_L|0)\rangle$ with $p_L^2=2$ and $(p_L|0)\in\Gamma$ are massless gauge vectors.  The first set generates an algebra $\mathfrak{u}_1^{\oplus\,n}$, the second set generates $\mathfrak{u}_1^{\oplus\,16+n}$, and the third set of states, if non-empty, may enhance the $\mathfrak{u}_1^{\oplus\,16+n}$ to a nonabelian gauge algebra.  The total rank of nonabelian and abelian gauge factors from the second and third sets should be $16+n$, and if there are no abelian factors (other than the $\mathfrak{u}_1^{\oplus\,n}$ from the first set of states), we say the gauge algebra is maximally enhanced.  The NS sector spacetime scalar states $\al^i_{-1}\wtpsi^a_{-\hlf}|0,(0|0)\rangle$ are massless moduli for the lattice compactification.  The R sector furnishes massless fermion states that are the superpartners of the bosons listed above.

In the case $n=0$, giving ten-dimensional heterotic theories, we need to start with an even self-dual rank sixteen lattice.  It is known that there are exactly two of these.  One of them is $\Gamma_{E_8}\oplus\Gamma_{E_8}$, two copies of the $E_8$ root lattice, and this gives rise to a $\mathfrak{e}_8\oplus\mathfrak{e}_8$ gauge algebra, and the other is $\Gamma_{D_{16}}+\langle(\hlf^{16})\rangle$, the $D_{16}$ root lattice plus a shift vector (i.e.~all vectors in $\R^{16}$ of the form $v+k\lp\hlf^{16}\rp$, where $v\in\G_{D_{16}}$ and $k\in\Z$), giving a $\mathfrak{d}_{16}$ gauge algebra \footnote{We give our conventions for the basis of the $ADE$ lattices in appendix \ref{app: ADE lattices}}. Neither case has any moduli (apart from the dilaton) because the index $a$ does not run over any values.

For $n>0$ there is a $n(n+16)$-dimensional space of torus moduli available, and the gauge algebra is some rank $16+n$ Lie algebra which may or may not be maximally enhanced, along with $\mathfrak{u}_1^{\oplus n}$.

\subsection{Lattice construction}\label{subsection: Lattice construction}
We will now present a method for constructing examples (particular points in the moduli space) of even self-dual lattices of signature $(16+n,n)$.  This construction, as we will see below, allows us to be able to choose the details in order to generate examples of non-supersymmetric heterotic strings which sit at extrema of the one-loop partition function. The construction described below is quite close to the construction used in \cite{Font_2020} and recently used in \cite{honda2026} to obtain the maximal enhancements for $T^2$ compactifications of $SO(16)\times SO(16)$ heterotic theory.

The most basic version of our construction works as follows.  Choose one of the two rank sixteen even self-dual Euclidean lattices, so either $\G_{16}=\G_{E_8}\oplus\G_{E_8}$ or $\G_{16}=\G_{D_{16}}+\langle(\hlf^{16})\rangle$, and then choose another lattice $\La$ which is an even rank $n$ Euclidean lattice.  Then define $\G_\La\subseteq\La^\vee\oplus\ov{\La^\vee}\subset\R^{n,n}$ by
\be
\G_\La=\left\{(p_L|p_R)\mid p_L,p_R\in\La^\vee,p_L-p_R\in\La\right\}.
\ee
Equivalent formulations are
\be
\G_\La=\left\{(u+y|y)\mid u\in\La,y\in\La^\vee\right\},\qquad\G_\La=\left\{(x|v+x)\mid x\in\La^\vee,v\in\La\right\}.
\ee
This lattice of course has signature $(n,n)$ by construction.  It is even, $(u+y)^2-y^2=u^2+2u\cdot y\in 2\Z$ for all $u\in\La$ and $y\in\La^\vee$ since $u\cdot y\in\Z$ and $u^2\in 2\Z$ since $\La$ is even.  It is also self-dual.  To show this, we simply need to show that $\G_\La^\vee\subseteq\G_\La$ (as $\G_\La\subseteq\G_\La^\vee$ already follows since $\G_\La$ is even).  If $(p_L|p_R)\in\G^{\vee}$, then we must have $p_L\cdot u\in\Z$ for all $u$. This implies $p_L\in\La^\vee$.  Similarly, we have $p_R\cdot v\in\Z$, which implies $p_R\in\La^\vee$. Lastly, it is easy to see that $(p_L-p_R)\cdot y=(p_L|p_R)\cdot(y|y)\in\Z$ for all $y$, which implies $p_L-p_R\in(\La^\vee)^\vee=\La$.  Thus, $\G_\La^\vee\subset\G_\La$, as claimed.  We can then construct the full $(16+n,n)$ lattice as $\G=\G_{16}\oplus\G_\La$, which will be even self-dual.

As an example, consider $n=1$.  The even Euclidean lattices of rank one are all of the form
\be
\La_N=\sqrt{2N}\Z,\qquad N\in\Z, N>0.
\ee
In particular, $N=1$ corresponds to the $A_1$ root lattice, $\La_1\cong\G_{A_1}$.  The dual lattices are
\be
\La_N^\vee=\frac{1}{\sqrt{2N}}\Z.
\ee
Then we have
\be
\G_{\La_N}=\left\{\lp\frac{k}{\sqrt{2N}}\Bigl\vert\frac{\ell}{\sqrt{2N}}\rp\Bigl\vert\;k,\ell\in\Z,k-\ell\equiv 0\ (\text{mod}\ 2N)\right\}
\ee
This can be written in a more familiar form as
\be
\G_{\La_N}=\left\{\lp\frac{1}{\sqrt{2}}\lp\frac{n}{\sqrt{N}}+w\sqrt{N}\rp\Bigl\vert\frac{1}{\sqrt{2}}\lp\frac{n}{\sqrt{N}}-w\sqrt{N})\rp\rp\Bigl\vert n,w\in\Z\right\},
\ee
the Narain momentum lattice for a single circle of radius $R=\sqrt{N}$ (in units where $\al'=1$)\footnote{If we instead used a construction
\be
\G_\La'=\left\{(p_L|p_R)\mid p_L,p_R\in\La^\vee,p_L+p_R\in\La\right\}=\left\{(u-y|y)\mid u\in\La,y\in\La^\vee\right\},
\ee
then $\G_{\La_N}'$ is the Narain momentum lattice for a circle of radius $R=1/\sqrt{N}$, which is physically equivalent (T-dual) to the construction above.}.

In particular, if $\La$ is an $ADE$ root lattice, then this construction gives a point in the Narain moduli space where the supersymmetric heterotic string has a maximally enhanced gauge algebra.  For instance if $\G_{16}=\G_{E_8}\oplus\G_{E_8}$ and $\La=\G_{A_4}$, then this is a point with $\mathfrak{e}_8\oplus\mathfrak{e}_8\oplus\mathfrak{a}_4$ gauge algebra on the left (and $\mathfrak{u}_1^{\oplus 4}$ on the right).

Now let us generalize this construction further.  We will construct a signature $(16+n,n)$ lattice $\Gamma$ by picking a rank $16+n$ Euclidean lattice $\La_L$, a rank $n$ Euclidean lattice $\La_R$, and a linear map $\rho:\R^n\rr\R^{16+n}$.  From these ingredients we define
\be\label{gamma element}
\Gamma=\left\{(v+\rho(u)|u)\mid v\in\Lambda_{L},u\in\Lambda_{R}^\vee\right\}
\ee
An element of $\Gamma$ has the norm squared
\be\label{general construction norm}
v^{2}+2v\cdot\rho(u)+\rho(u)^2-u^2
\ee
The squared norm in \eqref{general construction norm} should be even for all $v$ and $u$ for the lattice $\Gamma$ to be even. Therefore, $\La_L$ has to be even (which can be seen by setting $u=0$), and $\rho(u)^2-u^2$ should be an even integer for all $u\in\La_R^\vee$ (which can be seen by setting $v=0$). We then also need $v\cdot\rho(u)\in\Z$ for all $v\in\La_L$ and $u\in\La_R^\vee$, implying that $\rho(\La_R^\vee)\subseteq\La_L^\vee$.  These conditions are sufficient for $\G$ to be even.  We also need to find conditions for $\G$ to be self-dual. If $(k_L|k_R)\in\G^\vee$, then we must have $k_L\cdot v=(k_L|k_R)\cdot(v|0)\in\Z$, $\forall v\in\La_L$ which implies that $k_L\in\La_L^\vee$, and we must have \be
(\rho^T(k_L)-k_R)\cdot u=k_L\cdot\rho(u)-k_R\cdot u=(k_L|k_R)\cdot(\rho(u)|u)\in\Z\;\;\;\;\;\;\forall u\in\La_R^\vee
\ee
where $\rho^{T}:\mathbb{R}^{16+n}\rr \mathbb{R}^{n}$ defined such that
\be
\rho^{T}(k)\cdot u=k\cdot\rho(u)\;\;\;\;\;\;\forall k\in \mathbb{R}^{16+n},\;u\in\mathbb{R}^{n}
\ee
Therefore, $\rho^T(k_L)-k_R\in\La_R$. Thus, we can write
\be\label{Gamma dual presentation}
\G^\vee=\left\{(x|y+\rho^T(x))\mid x\in\La_L^\vee,y\in\La_R\right\}.
\ee
Self-duality then requires the following conditions
\begin{itemize}
\item[(i)] $\La_R$ is even because the elements of $\G^\vee$ should be even when $x=0$
\item[(ii)] $\rho^T(x)^2-x^2\in 2\Z$, $\forall x\in\La_L^\vee$ because the elements of $\G^\vee$ should be even when $y=0$
\item[(iii)] $\rho^T(\La_L^\vee)\subseteq\La_R^\vee$ to ensure that $(x|\rho^T(x))\in\G^\vee$ is also in $\G$
\item[(iv)] $\rho(\La_R)\subseteq\La_L$ (so that $(0|y)\in\G^\vee$ is also in $\G$)
\item[(v)] $\rho^T(\La_L)\subseteq\La_R$ (so that $(u|0)\in\G$ is also in $\G^\vee$)
\item[(vi)] $(1-\rho\rho^T)(\La_L^\vee)\subseteq\La_L$ and $(1-\rho^T\rho)(\La_R^\vee)\subseteq\La_R$
\end{itemize}
The first condition in (vi) can be derived by considering some $(p_L|p_R)\in\G$ and using the fact that $p_L-\rho(p_R)\in\La_L$. Applying this fact to $(x|\rho^T(x))\in\G^\vee$ requires $(1-\rho\rho^T)(\La_L^\vee)\subseteq\La_L$. The second condition can be derived by considering a general element of $\Gamma^\vee$ instead.

Not all of the conditions in (i)-(vi) on $\rho$ are independent.  Indeed, if we fix even lattices $\La_L$ and $\La_R$, then the condition $\rho(\La_R)\subseteq\La_L$ is equivalent to the condition $\rho^T(\La_L^\vee)\subseteq\La_R^\vee$, and similarly $\rho^T(\La_L)\subseteq\La_R$ is equivalent to $\rho(\La_R^\vee)\subseteq\La_L^\vee$.  Also, since the condition $(1-\rho^T\rho)(\La_R^\vee)\subseteq\La_R$ is equivalent to the statement that $w\cdot(1-\rho^T\rho)(u)\in\Z$ $\forall w,u\in\La_R^\vee$, we can show that it is actually implied by the condition $\rho(u)^2-u^2\in 2\Z$ by doing the following expansion
\begin{align}
w\cdot(1-\rho^T\rho)(u)&=w\cdot u-\rho(w)\cdot\rho(u)\nonumber\\
&=-\hlf\ls\lp \rho(w)^2-w^2\rp+\lp \rho(u)^2-u^2\rp-\lp\rho(u+w)^2-(u+w)^2\rp\rs\in\Z.
\end{align}
In the same way, $\rho^T(x)^2-x^2\in 2\Z$ implies $(1-\rho\rho^T)(\La_L^\vee)\subseteq\La_L$. It can be shown that for self-dual $\Gamma$, $\rho^{T}(x)^2-x^2\in2\Z\forall x\in\Lambda^\vee_L$ is equivalent to $\rho(u)^2-u^2\in2\Z\forall u\in\Lambda^\vee_R$. Since $\Gamma$ is self-dual, any $x\in\Lambda^\vee_L$ can be written as $v+\rho(u)$ with $v\in\Lambda_L$ and $u\in\Lambda_R^\vee$. Then we can do the following manipulation
\begin{align}
\rho^T(x)^2-x^2=&\rho^T(v+\rho(u))^2-(v+\rho(u))^2\nonumber\\
=&\rho^T(v)^2-v^2+2\rho^T(v)\cdot(u+y)+(u+y)\cdot(u+y)\nonumber\\
\label{rhoT combination}
&-2v\cdot\rho(u)-\rho(u)\cdot\rho(u)
\end{align}
where $y\in \Lambda_R$. We used the fact that $(1-\rho^T\rho)(\Lambda_R^\vee)\subseteq \Lambda_R$ (which is a consequence of $\rho(u)^2-u^2\in2\Z$) to write $\rho^T\rho(u)$ as $u+y$. Now, using the fact that $\rho^T(\Lambda_L)\subseteq \Lambda_R,\;\rho(\Lambda_R)\subseteq \Lambda_L$, the fact that $\Lambda_L$ and $\Lambda_R$ are even lattices, and the definition of $\rho^T$, it can be shown that the expression in \eqref{rhoT combination} reduces to
$$
\rho^T(x)^2-x^2=u^2-\rho(u)^2\;\;(\text{mod }2)
$$
but since $\rho(u)^2-u^2\in2\Z$, we have $\rho^T(x)^2-x^2\in2\Z$. Therefore, the only independent conditions on $\rho$ are
\be\label{conditions on rho}
\rho(u)^2-u^2\in2\Z\;\;\forall u\in\Lambda^\vee_R,\efill \rho(\Lambda_R)\subseteq \Lambda_L,\efill \rho(\Lambda^\vee_R)\subseteq \Lambda^\vee_L
\ee
If we use the last condition to define $\rho$ as a map on $\Lambda^\vee_R$ only, then there are only two independent conditions on $\rho$.

Let us also note that if $\rho$ is a map satisfying all of the required conditions, then $\rho'=\rho+UV^T$, where $U\in\La_L$ and $V\in\La_R$, will also satisfy all the conditions and will construct the same lattice $\G$.

Now suppose we have $\La_L$, $\La_R$ and $\rho$ satisfying these conditions, and we use them to construct an even self-dual $\G$.  Let $A_L=\La_L^\vee/\La_L$ and $A_R=\La_R^\vee/\La_R$ be the discriminant groups of $\La_L$ and $\La_R$.  We can define a group homomorphism $\tilde{\rho}:A_R\rr A_L$ by
\be
\tilde{\rho}([u]):=[\rho(u)].
\ee
This is well-defined since for $u\in\La_R^\vee$, $v\in\La_R$, we have $[\rho(u+v)]=[\rho(u)+\rho(v)]=[\rho(u)]$ since $\rho(\La_R)\subseteq\La_L$.  We can show that $\tilde{\rho}$ is surjective: given $x\in\La_L^\vee$, we have $x=\rho\rho^T(x)+(1-\rho\rho^T)(x)$, and then $[x]=[\rho\rho^T(x)]=\tilde{\rho}([\rho^T(x)])$.  If $[u]$ is in the kernel of $\tilde{\rho}$, $[0]=\tilde{\rho}([u])=[\rho(u)]$, then $\rho(u)\in\La_L$ and hence $u\cdot(w-(1-\rho^T\rho)(w))=\rho(w)\cdot\rho(u)\in\Z$, $\forall w\in\La_R^\vee$.  But since $(1-\rho^T\rho)(\La_R^\vee)\subseteq\La_R$, this implies $u\cdot w\in\Z$, $\forall w\in\La_R^\vee$ and hence $u\in\La_R$ and $[u]=[0]$.  Thus $\tilde{\rho}$ is also injective and thus $\tilde{\rho}$ is an isomorphism between $A_R$ and $A_L$.  The inverse map is $\tilde{\rho}^{-1}([x]):=[\rho^T(x)]$.

To summarize, we pick even Euclidean lattices $\La_L$ and $\La_R$ of rank $16+n$ and $n$ respectively, whose discriminant groups $A_L$ and $A_R$ are isomorphic, and we try to find a linear map $\rho:\La_R^\vee\rr\La_L^\vee$ satisfying $\rho(u)^2-u^2\in 2\Z$, $\forall u\in\La_R^\vee$ and which induces an isomorphism $\tilde{\rho}:A_R\rr A_L$ (it may not always be possible to find such a $\rho$).  Then the other required conditions on $\rho$ follow (for instance $\tilde{\rho}$ being injective implies $\rho(\La_R)\subseteq\La_L$, and the surjectivity of $\tilde{\rho}$ can be used to show that $\rho^T(x)^2-x^2\in 2\Z$, $\forall x\in\La_L^\vee$) and the lattice $\G$ given in~\eqref{gamma element} is even self-dual.

We now point out the connections between our construction and the construction used in \cite{Font_2020}. Our $\Lambda_R$ is what the authors of \cite{Font_2020} call $T$ (both of these lattices are even lattices). If we restrict ourselves to the points of maximal enhancement, then what we call $\Lambda_L$ is the overlattice $M$ in \cite{Font_2020}. In addition, the lattice $L$ in \cite{Font_2020} corresponds to the root lattice inside $\Lambda_L$. For non-maximal enhancement points, $\Lambda_{L}$ isn't necessarily the overlattice of the root lattice (in some cases, the root lattice and $\Lambda_L$ may not even have the same rank). Finally, quadratic forms $q_{L}$ and $q_{R}$ required in \cite{Font_2020} have a nice interpretation in our language. A quadratic form $q$ is defined over the discriminant group of an even lattice $\Lambda^\vee/\Lambda$ as
\be
q([x])=x^2\text{ mod }2\text{, where }x\in\Lambda^\vee
\ee
The authors of \cite{Font_2020} require the left and right quadratic forms, i.e., $q_{L}$ and $q_{R}$ (defined on the discriminant groups of $M$ and $T$ lattices) to be congruent. In our language, this translates to the quadratic forms over $A_L$ and $A_R$ being congruent. These forms can be calculated to give
\begin{align}
q_{L}([v+\rho(u)])=q_{L}([\rho(u)])&=\rho(u)^2\;\;\text{mod }2\;\;\;\text{ where }v\in\Lambda_L,\;u\in\Lambda^{\vee}_R\\
q_{R}([u])&=u^2\;\;\text{mod }2\;\;\;\text{ where }u\in\Lambda^{\vee}_R
\end{align}
where we used the fact that any $x\in \Lambda^\vee_L$ can be written as $v+\rho(u)$ with $v\in\Lambda_L$ and $u\in\Lambda^\vee_R$, which follows from the self duality of $\G$. Therefore, the congruency of the left and right quadratic forms nicely translates to $\rho(u)^2-u^2\in2\Z$, which is already a required condition in our approach. If we use the parametrization in \eqref{Gamma dual presentation}, then the congruency of quadratic forms would imply $\rho^{T}(x)^2-x^2\in 2\Z\;\forall x\in\Lambda^{\vee}_{L}$, which is again a required (although not independent) condition in our approach.
\subsection{Non-supersymmetric heterotic orbifolds}
The next step is to gauge a global $\Z_2$ symmetry of the worldsheet theory of the compactified supersymmetric heterotic string in such a way to break the spacetime supersymmetry (the worldsheet supersymmetry will be preserved).  The symmetry will be generated by a group $g_\d$ which acts on states labeled by $p\in\Gamma$ as the phase $\exp[2\pi i\,\d\cdot p](-1)^{F_s}$, where $\d$ is a vector in $\R^{16+n,n}$ and $(-1)^{F_s}$ simply acts as $+1$ in the NS sector and $-1$ in the R sector.  In order for this generator to be order two, we must have $2\d\in\Gamma$. Also, it turns out that we must have $\d^2\in\Z$ or else the resulting $\Z_2$ symmetry is anomalous and cannot be gauged (the obstacle to gauging manifests as a failure of modular invariance in the orbifold partition function) \cite{Alvarez-Gaume:1986ghj, Ginsparg:1986wr}.  Note that shifting $\d$ by a lattice vector does not change the action of $g_\d$, so the construction really depends on an equivalence class in the coset $\hlf\G/\G$.  Moreover, acting on $\d$ with a lattice automorphism (preserving inner products) gives a physically equivalent construction.

After gauging, we have spectrum contributions from both the untwisted and twisted sectors, with NS and R sectors in each case.  Level matching is still governed by~\eqref{eq:LevelMatching}, but in the untwisted sector, we need to restrict to $g_\d$-invariant states, and in the twisted sector, states are labeled by a vector $p\in\G+\d=\{v+\d\mid v\in\G\}$, the GSO projection is flipped, and we again have to project onto invariant states. In more detail, sector by sector, the low-lying states are:
\begin{itemize}
\item In the untwisted NS sector we have the same metric, $B$-field, and dilaton states $\al^\m_{-1}\wtpsi^\n_{-\hlf}|q,(0|0)\rangle$, massless vectors $\al^\m_{-1}\wtpsi^a_{-\hlf}|q,(0|0)\rangle$, $\al^i_{-1}\wtpsi^\m_{-\hlf}|q,(0|0)\rangle$, and $\wtpsi^\m_{-\hlf}|q,(p_L|0)\rangle$ where $p_L^2=2$ and $\d\cdot(p_L|0)\in\Z$, and massless scalars $\al^i_{-1}\wtpsi^a_{-\hlf}|q,(0|0)\rangle$.  The vectors lead to a rank $16+n$ gauge algebra as before, but reduced from the supersymmetric algebra in general.  The scalars will be called pseudo-moduli for reasons we'll discuss below.
\item In the twisted NS sector, states are labeled by a vector $(p_L|p_R)\in\G+\d$ and with an even number of $\wtpsi$ raising operators.  Invariance under $g_\d$ requires either $\d\cdot(p_L|p_R)\in\Z+\hlf$ if $\d^2\in 2\Z$, or $\d\cdot(p_L|p_R)\in\Z$ if $\d^2\in 2\Z+1$.  States with $N_R>0$ will be massive.  States with $N_R=0$ will be massive if $p_R^2>1$, massless if $p_R^2=1$, and tachyonic if $p_R^2<1$.  For the tachyonic states, level matching requires $N_L=0$ and $p_L^2=p_R^2+1$.  For the $p_R^2=1$ massless states we either have massless scalars with $N_L=0$ and $p_L^2=2$, or we have $N_L=1$ and $p_L^2=0$ giving scalars $\al^i_{-1}|q,(0|p_R)\rangle$ and vectors $\al^\m_{-1}|q,(0|p_R)\rangle$.  The vectors enhance the ``right-moving'' $\mathfrak{u}_1^{\oplus\,n}$ part of the gauge algebra to a non-abelian rank $n$ algebra.  Massless scalars of the first type ($p_L^2=2$) can always become tachyonic under infinitesimal deformation of the lattice and lead to ``knife-edges''\cite{Ginsparg:1986wr, Fraiman:2023cpa}, while the scalars of the second type ($p_L^2=0$) become massive under deformation and combine with (get eaten by) the massless vectors to form massive vectors in the deformed theory.
\item The untwisted R sector gives massless or massive fermionic states.  Massless states only exist when we have $(p_L|0)\in\G$ with $p_L^2=2$ and $\d\cdot(p_L|0)\in\Z+\hlf$.
\item The twisted R sector also contains only massless or massive fermionic states, with massless states requiring $(p_L|0)\in\G+\d$ with $p_L^2=2$ and $\d\cdot(p_L|0)\in\Z$ if $\d^2\in 2\Z$ or $\d\cdot(p_L|0)\in\Z+\hlf$ if $\d^2\in 2\Z+1$.
\end{itemize}
Spacetime supersymmetry is broken in this theory, as evidenced by the lack of gravitini and the mismatch between bosonic and fermionic spectra.
\subsection{One-loop effective potential}\label{subsection: One-loop potential}
Due to the lack of spacetime supersymmetry, the one-loop partition function does not vanish, and can lead to an effective potential in $(10-n)$ dimensions.  The partition function is given by
\be\label{one loop partition function}
Z(\tau,\bar{\tau})=\tau_2^{-\frac{8-n}{2}}\eta^{-24}\bar{\eta}^{-8}\left[\ov{V_8}\s_V-\ov{S_8}\s_S-\ov{C_8}\s_C+\ov{O_8}\s_O\right],
\ee
where expressions for $\ov{V_8}$, $\ov{S_8}$, $\ov{C_8}$, and $\ov{O_8}$ are given in Appendix~\ref{app: theta functions}, and where the $\s$'s represent lattice sums\footnote{The appearance of $\d^2$ in the parity condition for the $\s_C$ and $\s_O$ sums is required by modular invariance, and is consistent with the spectrum discussed above.},
\begin{equation*}
\s_V=\sum_{\substack{(p_L|p_R)\in\G \\ \d\cdot(p_L|p_R)\in\Z}}q^{\hlf p_L^2}\bar{q}^{\hlf p_R^2},\qquad\s_S=\sum_{\substack{(p_L|p_R)\in\G \\ \d\cdot(p_L|p_R)\in\Z+\hlf}}q^{\hlf p_L^2}\bar{q}^{\hlf p_R^2},
\end{equation*}
\be
\s_C=\sum_{\substack{(p_L|p_R)\in\G+\d \\ \d\cdot(p_L|p_R)\in\Z+\hlf\d^2}}q^{\hlf p_L^2}\bar{q}^{\hlf p_R^2},\qquad\s_O=\sum_{\substack{(p_L|p_R)\in\G+\d \\ \d\cdot(p_L|p_R)\in\Z+\hlf(1+\d^2)}}q^{\hlf p_L^2}\bar{q}^{\hlf p_R^2}.
\ee

Nominally, the contribution to the effective potential (in string frame) is
\be\label{potential for Tn}
V=-\lp 4\pi^2\al'\rp^{-\frac{10-n}{2}}\int_{\mathcal{F}}\frac{d^2\tau}{2\tau_2^2}Z(\tau,\bar{\tau}),
\ee
where $\mathcal{F}$ is the usual $\SL(2,\Z)$ fundamental domain in the $\tau$ upper half-plane.  Of course, this is only sensible if this integral is convergent, which happens if and only if there are no tachyons in the low-lying spectrum.  If there are tachyons, then there will be a divergent contribution coming from the $\ov{O_8}\s_O$ term.

When $n>0$, giving VEVs to the pseudo-moduli fields changes the lattice $\G$, and hence changes the potential $V$.  A common parameterization of the moduli which is especially useful when thinking of the lattice as representing a $T^n$ compactification of one of the ten-dimensional heterotic theories, is to give the metric $G$ and $B$-field on the torus, along with $16n$ Wilson line moduli from the sixteen gauge fields in the Cartan threading each of the $n$ circles on the torus.

However, if we want to expand around a general point in the lattice moduli space corresponding to an even self-dual lattice $\G$, there is a more convenient parameterization of the moduli, which is simply in terms of a $(16+n)\times n$ matrix $M_{ia}$.  Under such a deformation, we replace each vector $(p_L|p_R)\in\G$ by a vector $(p_L'|p_R')$ in $\G'\subset\R^{16+n,n}$ given by
\be\label{momentum deformtation}
p_L'=\sqrt{\mathbb{I}_{16+n}+MM^T}\,p_L+Mp_R,\qquad p_R'=M^Tp_L+\sqrt{\mathbb{I}_n+M^TM}\,p_R.
\ee
The deformed lattice $\G'$ is the full set of such vectors.  Of course, since the involution vector needs to be order two in this new lattice, we must deform it as well,
\be
\d=(\d_L|\d_R)\mapsto\d'=(\d_L'|\d_R')=(\sqrt{\mathbb{I}_{16+n}+MM^T}\,\d_L+M\d_R\;|\;M^T\d_L+\sqrt{\mathbb{I}_n+M^TM}\,\d_R).
\ee

\subsection{Finding extrema}

Now we can examine the condition that the potential $V$ is an extremum of the pseudo-moduli, i.e.~that $\frac{\p}{\p M_{ia}}V|_{M=0}=0$ for all $i=1,\cdots,16+n$ and $a=1,\cdots,n$.  Using the fact that
\be
\left.\frac{\p}{\p M_{ia}}p_L^2\right|_{M=0}=\left.\frac{\p}{\p M_{ia}}p_R^2\right|_{M=0}=2p_{L\,i}p_{R\,a},
\ee
we have
\be\label{sigma_V der}
S_{V\,ia}:=\left.\frac{\p\s_V}{\p M_{ia}}\right|_{M=0}=-4\pi\tau_2\sum_{\substack{(p_L|p_R)\in\G \\ \d\cdot(p_L|p_R)\in\Z}}p_{L,i}p_{R,a}q^{\hlf p_L^2}\bar{q}^{\hlf p_R^2},
\ee
and similar expressions hold for derivatives of $\s_{S},\s_{O}$ and $\s_{C}$, which we will denote $S_{S\,ia}$, $S_{C\,ia}$, and $S_{O\,ia}$ respectively.  In order for us to have an extremum of the effective potential (with respect to the pseudo-moduli fields) all four of these quantities should vanish for all components $i$ and $a$ and all values of $\tau$\footnote{In fact this is a sufficient condition but may not be necessary.  For one, theta function identities imply that $\ov{V_8}=\ov{S_8}=\ov{C_8}$, so it would be sufficient for the combinations $S_{O\,ia}$ and $S_{V\,ia}-S_{S\,ia}-S_{C\,ia}$ to vanish.  Generally speaking, though, the $q$ series appearing in $\s_V$, $\s_S$, and $\s_C$ are different, with different exponents and multiplicities, and so each of those three terms must vanish separately.  Another comment is that we actually don't need these quantities to vanish as functions of $\tau$; it would be enough for them to vanish after integrating over $\tau$.  However, extrema of this sort seem somewhat unlikely, or at least hard to find.}.  Equivalently, given a vector $v^a\in\R^n$, we can define $S_{V\,i}(v)=v^aS_{V\,i}$, and then we get an extremum if we can find a basis of vectors $v$ for which $S_{V\,i}(v)=0$ (and similarly for $S$, $C$, and $O$).

The key to finding pairs $(\G,\d)$ which extremize the potential is to use symmetries.  In particular, let's consider the construction of Section~\ref{subsection: Lattice construction} above for the special case where $\La_R$ is an ADE root lattice and $\d=(\d_L|0)$.  Clearly there is a basis of $\R^n$ consisting of root vectors.  We will show that $S_{A\,i}(\al)=0$ for all $i$ and $A=V,S,C,O$ and roots $\al$.  Indeed recall that Weyl reflection in $\al$ acts on vectors $v\in\R^n$ as
\be
w_\al(v)=v-(\al\cdot v)\al,
\ee
and satisfies that $w_\al(v)^2=v^2$ and that $\al\cdot w_\al(v)=-\al\cdot v$.  If $y\in\La_R^\vee$, then also $w_\al(y)\in\La_R^\vee$.  Moreover, for $(u+\rho(y)|y)\in\G$, we can find $u'=u+\rho(y)-\rho(w_\al(y))=u+(\al\cdot y)\rho(\al)\in\La_L$, so the vector $(u+\rho(y)|w_\al(y))=(u'+\rho(w_\al(y))|w_\al(y))\in\G$.  In other words, we can lift $w_\al$ to an involution $\tilde{w}_\al$ on $\G$ defined by
\be
\tilde{w}_\al((p_L|p_R))=(p_L|w_\al(p_R)).
\ee
This involution preserves delta parity, $\d\cdot\tilde{w}_\al((p_L|p_R))=\d\cdot(p_L|p_R)$ (since $p_L$ doesn't change and $\d_R=0$) but it flips the sign of the summand in $S_{V\,i}(\al)$ or $S_{S\,i}(\al)$, e.g.~
\be
-\frac{S_{V\,i}(\al)}{4\pi\tau_2}=\sum_{\substack{(p_L|p_R)\in\G \\ \d\cdot(p_L|p_R)\in\Z}}p_{L\,i}(\al\cdot p_R)q^{\hlf p_L^2}\bar{q}^{\hlf p_R^2}=\sum_{\substack{(p_L|p_R)\in\G \\ \d\cdot(p_L|p_R)\in\Z}}p_{L\,i}(\al\cdot w_\al(p_R))q^{\hlf p_L^2}\bar{q}^{\hlf w_\al(p_R)^2}=\frac{S_{V\,i}(\al)}{4\pi\tau_2},
\ee
and hence $S_{V\,i}(\al)=0$ and similarly for $S_{S\,i}(\al)$.  The action of $\tilde{w}_\al$ can also be extended to $\G+\d$ with the same properties (if $(p_L|p_R)\in\G+\d$, then $\tilde{w}_\al((p_L|p_R))=(p_L|w_\al(p_R))\in\G+\d$ as well, $\d\cdot\tilde{w}_\al((p_L|p_R))=\d\cdot(p_L|p_R)$, and $\al\cdot w_\al(p_R)=-\al\cdot p_R$), and so can be used to show that $S_{C\,i}(\al)=S_{O\,i}(\al)=0$.

In summary, if we take $\La_R$ to be a rank $n$ ADE root lattice, find another even euclidean lattice $\La_L$ of rank $16+n$ such that we have a quadratic form-preserving isomorphism between discriminant groups $\La_R^\vee/\La_R$ and $\La_L^\vee/\La_L$, and pick a $\d_L\in\hlf\La_L$, $\d_L^2\in\Z$, then the corresponding $\G$ and $\d=(\d_L|0)$ will give rise to an extremum of the nonsupersymmetric heterotic string on $T^n$\footnote{More precisely, perhaps, it will give us a point in the moduli space of the supersymmetric heterotic string on $T^n$, and a $\Z_2$ supersymmetry-breaking orbifold at that point, such that the resulting orbifold theory's one-loop potential is extremized.}.

One may ask whether the conditions can be relaxed.  As long as $\al$ is any root (i.e.~$\al^2=2$) of $\La_R$, then $w_\al$ will act as an involution of $\La_R^\vee$ and can be lifted to an involution $\tilde{w}_\al$ of $\G$.  In order to use such involutions to prove that all first derivatives of $\s_V$ vanish, we must be able to find a basis of such roots.  In other words, we require that the root lattice of $\La_R$ is maximal rank.  Fortunately for our purposes, when $n<8$ this only happens when $\La_R$ is itself an ADE root lattice.  We can also consider relaxing the condition that $\d_R=0$.  Checking the requirements, we find that the precise condition is that $\d_R\cdot\al\in\Z$ for every $\al$ we want to use for a Weyl reflection.  This means that we need the roots with integer inner products to again generate a sublattice of maximum rank.  If every root of $\La_R$ has integer inner product with $\d_R$, then it means that $\d_R\in\La_R^\vee$, and we can therefore shift by an element of $\G$ to set $\d_R=0$.  So we may assume that the surviving roots generate a proper maximum sublattice.  For any $\La_R$ which is a direct sum of $A$-type root lattices (which is the only possibility for $n<4$), this can't happen since an $A_m$ root lattice has no maximal rank proper root system.  For $n=4$ this can happen with the $D_4$ lattice (taking $\d_R=(\hlf^2,0^2)$ for instance preserves a $4A_1$ root system), but we will not include such solutions in this paper.  At higher ranks this will happen commonly.

There was nothing in this construction that required us to use Weyl reflections on the right to prove extremality.  We can repeat the arguments using Weyl reflections of $\La_L$.  Specifically, if we can find a rank $16+n$ even euclidean lattice $\La_L$ whose root lattice is maximal (for these ranks this does not imply that $\La_L$ is itself an ADE root lattice), an even euclidean rank $n$ lattice $\La_R$ such that the discriminant groups $\La_L^\vee/La_L$ and $\La_R^\vee/\La_R$ enjoy a quadratic form-preserving isomorphism, and a vector $\d=(\d_L|\d_R)\in\hlf\G$ such that $\d^2\in\Z$ and we can find a basis of $\R^{16+n}$ consisting of roots $\al$ in $\La_L$ with $\d_L\cdot\al\in\Z$, then we will get an extremum.  In this case, looking at the spectrum of the resulting theory, we see that the left-moving gauge algebra will be a maximal (i.e.~rank $16+n$) enhancement.  These points of maximal enhancement thus give rise to extrema of the one-loop potential.  These have been carefully studied for $n=1$ by~\cite{Fraiman:2023cpa} and recently for $n=2$ by~\cite{honda2026}.  Our approach gives an alternative pathway to deriving those same results.  However, in this paper we focus on the case where we use Weyl reflections in $\La_R$ instead, and this will generically lead to extremal points that do not have maximal gauge enhancement.

\section{Finding extrema with $\Lambda_{R}$ being $ADE$}\label{sec 3: Finding extrema}

Our strategy in this paper will repeatedly be to construct an even self-dual $(16+n,n)$ signature lattice with $\Lambda_R$ being an $ADE$ lattice, so that we will have an extremum of the pseudo-moduli potential given in the previous section. In order to construct these $(16+n,n)$ signature lattices, we want to find $\Lambda_L$s that can appear for a given $ADE$ lattice $\Lambda_R$. Somewhat surprisingly, it turns out that known algorithms will let us completely classify the possible $\La_L$ for fixed $\La_R$ - it is equivalent to enumerating all (isometry classes of) lattices within a given genus, where two lattices are in the same genus if they are equivalent upon tensoring with $\R$ or with $\Z_p$ for any prime $p$.  After classifying all possible $\La_R$, we then classify the possible involution vectors $\d$ that can be used to perform the supersymmetry-breaking orbifold. As we saw earlier, setting $\d_R=0$ is sufficient (and for $n\le 3$ necessary, without loss of generality) to ensure that the sums $\s_V,\s_S,\s_C$ and $\s_O$ are extrema with respect to the pseudo-moduli, and thus, we will look for involutions with $\d_R=0$. Since the root lattices of many of the $\Lambda_L$ that appear through this strategy have ranks less than $16+n$, this will lead to many examples of extrema with non-maximal enhancements. In contrast, the set of maximal enhancements for compactifications on $S^1$ \cite{Fraiman:2023cpa} and $T^2$ \cite{honda2026} are known; we recover some of them with our construction, but not all. For each $\d$, we will find the enhancement (maximal or non-maximal), and the tachyonic spectrum of states (if non-empty) with special interest in the tachyon-free examples. We provide the results for $\d$s, enhancements, and tachyonic spectra for the $n=1$ case in Table \ref{tab: rank 1 A1}. The results for $n=2,3$ and $4$ are provided in the ancillary Excel files for this paper. One additional Excel file provides the list of tachyon-free points.

A few comments about knife edges are in order. Knife edges occur when we have scalars with momentum $(p_L|p_R)$ in $\Gamma+\d$, satisfying $p^2_R=1$ and $p^2_L=2$ and $\d\cdot(p_L|p_R)\in\Z+\hlf(\d^2+1)$. In our construction, $p_R\in\Lambda_R^{\vee}+\d_R$, but to get an extremum, we are setting $\d_R=0$. Hence, $p_R$ lives in $\Lambda^\vee_R$, and therefore, if we take $\Lambda_R$ to be an $ADE$ lattice, then it is easy to see that those choices of $ADE$ lattices for $\Lambda_R$ whose duals don't have a vector with squared norm equal to $1$ will be free of knife edges. To see which choices fall in this category, let's start from irreducible $ADE$ lattices. From these lattices, we only have $A_3$ and $D_n\;(n\geq 4)$ whose duals can have a vector of unit squared norm. Now, if we consider reducible $ADE$ lattices (i.e. a direct sum of irreducible $ADE$ lattices), it suffices to look at the vectors with the smallest non-zero squared norm in the duals of all irreducible $ADE$ lattices. The results are tabulated in Table \ref{tab: smallest norm squared}. It is easy to see that $E_6,E_7$ and $E_8$ can't contribute to having a vector with norm squared of $1$. Moreover, we see that for $A_n$ with $n\geq 2$, the lowest squared norm is bigger than or equal to $2/3$ and thus, any two $A_n$s will produce a vector with a squared norm of at least $4/3$. This leaves $A_1$. We see that $A_1$ can't combine with any other $ADE$ lattice to give a vector of unit squared norm except another $A_1$. Therefore, there are only three ways to get a vector of unit squared norm , i.e., have an $A_3,\;D_n\;(n\geq 4),$ or $2A_1$ sublattice. Using this result, we see that the following $\Lambda_R$s are certainly free of knife edges for compactifications on $T^n$ with different $n$
\begin{itemize}
    \item $n=1:\;\Lambda_R=A_1$
    \item $n=2:\;\Lambda_R=A_2$
    \item $n=3:\;\Lambda_R=A_2+A_1$
    \item $n=4:\;\Lambda_R=A_4,\;2A_2$
\end{itemize}
For other $\Lambda_R$s, one needs to check that no vector $(p_L|p_R)$ with $p^2_L=2$ and $p_R^2=1$ exists in $\G+\d$ to show that a particular extremum is not a knife edge. This statement depends on $\Lambda_L$ and thus would need to be checked for individual $\Lambda_L$s.
\begin{table}[]
    \centering
    \begin{tabular}{|c|c|}
         \hline
         $ADE$ Lattice $(\Lambda)$&Smallest non-zero squared norm in $\Lambda^\vee$\\\hline
         $A_n\;(n\geq 1)$& $\frac{n}{n+1}$  \\
         $D_n\;(n\geq 4)$& $1$ \\
         $E_6$& $4/3$ \\
         $E_7$& $3/2$ \\
         $E_8$& $2$ \\\hline
    \end{tabular}
    \caption{Smallest non-zero squared norms for $ADE$ lattices}
    \label{tab: smallest norm squared}
\end{table}

\subsection{Finding $\Lambda_L$s}

To find $\Lambda_L$s, we proceed as follows. Fix an $ADE$ lattice $\Lambda_R$, say of rank $n$. As explained previously, the lattices $\Lambda_L$ we want satisfy the condition that the discriminant groups $A_L$ and $A_R$ are isomorphic, and the respective discriminant forms $q_L$ and $q_R$ are congruent. These two conditions define a unique \emph{genus} of Euclidean lattices of rank $16+n$. A genus is a collection of lattices which are pairwise isometric after tensoring with $\mathbb{R}$ and $\mathbb{Z}_p$, for every prime number $p$. Equivalently, we can see it as the collection of all lattices of given signature, with the same discriminant group and congruent discriminant forms. We let $\mathcal{G}_L$ be the genus of the lattices $\Lambda_L$ we look for. According to Kneser (\cite{Kneser:2002}, Satz~(21.3)), any given genus consists of finitely many lattices, up to isometry. Kneser also shows that there is an algorithm to enumerate all such lattices; in our case, we can therefore explicitly describe all isometry classes of lattices in $\mathcal{G}_L$. We briefly review the main ideas of the algorithm in our situation.

Let $\Lambda_L$ be any Euclidean lattice in $\mathcal{G}_L$; for instance, we may choose $\Lambda_L := 2E_8+\Lambda_R$. Let moreover $p$ be an odd prime number not dividing the order of the discriminant group $A_L$ of $\Lambda_L$. For every vector $x\in \Lambda_L\setminus p\Lambda_L$ of squared length divisible by $2p^2$, the Euclidean lattice
\[\Lambda_L^p(x) := \{y\in L\mid x\cdot y\in p\mathbb{Z}\} + \frac{\mathbb{Z}}{p}x\]
is also in the genus $\mathcal{G}_L$. The lattice $\Lambda_L^p(x)$ is called a \emph{$p$-neighbor} of $\Lambda_L$, and it satisfies
\[\Lambda_L/(\Lambda_L\cap \Lambda_L^p(x))\cong \Lambda_L^p(x)/(\Lambda_L\cap \Lambda_L^p(x))\cong \mathbb{Z}/p\mathbb{Z}.\]
The idea of Kneser's algorithm is, for a fixed prime number $p$, to iteratively enumerate all $p$-neighbors of lattices in $\mathcal{G}_L$ until one constructs a lattice for each of the finitely many isometry classes in $\mathcal{G}_L$.\footnote{In practice, one constructs only finitely many such $p$-neighbors at every iteration. As explained in \cite{Kneser:2002}, Hilfssatz (28.7), the isometry class of $\Lambda_L^p(x)$ is determined by the orbit of $[x]\in \Lambda_L/p\Lambda_L$ under the action of the group of isometries $O(\Lambda_L)$ of $\Lambda_L$.} If one chooses the prime number $p$ correctly, then this procedure eventually terminates, as guaranteed by the so-called \emph{Smith--Minkowski--Siegel mass formula}. We refer to \cite{Voight:2023} for a recent survey on Kneser's method of neighbors. This approach is a standard routine, and Kneser's algorithm has been implemented in most computer algebra systems supporting dedicated lattice computations. For our purpose, we resort to the software OSCAR \cite{OSCAR-book} and its implementation, which features some enhancements based on isometry enumeration\footnote{This implementation is mainly due to S.~Brandhorst, T.~Hofmann and S.~Muller, and it is open-source \cite{OSCAR}}. For the sake of the example, let us choose $\Lambda_R = A_3$. In that case, up to equivalence of lattices, there are exactly 16 Euclidean lattices in $\mathcal{G}_L$, such as the $ADE$ root lattices $2E_8 + A_3$ and $D_{19}$. 

Finally, for any given lattice $\Lambda_L$ in $\mathcal{G}_L$, determining the root sublattice of $\Lambda_L$, i.e., the largest $ADE$ lattice it contains, can be dealt with using an algorithm of Fincke--Pohst (see \cite{Cohen:1993}, Section~2.7.3). Roughly speaking, the idea of this algorithm is to find a large enough real hyperellipsoid centered around the origin of $\Lambda_L$ and containing all vectors of bounded length. Since such a hyperellipsoid contains only finitely many vectors of $\Lambda_L$, one can enumerate all of them, compute their lengths, and keep the ones of interest. For instance, one can apply this algorithm to find the coordinates for all the finitely many vectors of $\Lambda_L$ with a given length. In particular, by collecting all vectors of squared length 2 and by analyzing how they pair to one another, it is effectively possible to systematically recognize the root system of the Euclidean lattice $\Lambda_L$. In our previous example, where $\Lambda_R = A_3$, we find that all the 16 possibilities for $\Lambda_L$ have pairwise distinct root systems, all of rank 18 or 19, tabulated in Table \ref{tab: rank 3 A3}. We will write our $\Lambda_L$ as the direct sum of a root lattice $\Lambda_{\text{root}}$, with an ``extra lattice" $\Lambda_{\text{extra}}$, and add in the span of $k$ shift vectors $s_1,\cdots,s_k$ i.e.
\be
\Lambda_L=\lp\Lambda_{\text{root}}\oplus\Lambda_{\text{extra}}\rp+\langle \{s_1,\cdots,s_k\}\rangle=\left\{x+y|x\in\Lambda_{\text{root}}\oplus\Lambda_{\text{extra}},\ y\in\langle \{s_1,\cdots,s_k\}\rangle\right\}.
\ee
This extra lattice is required whenever we are working with $T^n$ compactifications and the rank of the root lattice $r$ is less than $16+n$. This means that the extra lattice is a $16+n-r$ dimensional lattice, which is even (with squared norms of $4$ or higher to avoid having any additional roots). Moreover, for each shift vector $s_\al$ there is an integer $N_\al>2$ such that $N_\al s_\al\in\Lambda_{\text{root}}\oplus \Lambda_{\text{extra}}$.
\subsection{Redundant $\d$s}
There can be redundancies in the list of $\d$s associated with repeated root-system components (e.g., $2A_9$ has two repeated $ A_9$s). To remove these redundancies, we consider permutation transformations of the root lattice (that we call $S$). Each of these transformations can be seen as a transformation on the underlying Euclidean space which permutes the coordinates corresponding to two or more identical $ADE$ components of the root system, and acts trivially on the additional lattice directions. These transformations can be represented by an orthogonal matrix. Not all the permutations are symmetries of the full $\Gamma$, and we will consider a permutation to be a symmetry only if it preserves the full $\Gamma$. To determine which permutations work, we need to consider transformations of the right lattice as well.

Consider a transformation $(S, T)$ on $\Gamma$ where $S$ acts on the left momentum and $T$ acts on the right momentum, and they map zero vectors to zero vectors. Since purely left and right moving sublattices of $\Gamma$ are $\Lambda_L\oplus 0$ and $0\oplus\Lambda_R$ respectively, the preservation of $\Gamma$ under the $(S,T)$ transformation requires $S(\Lambda_L)=\Lambda_L$ and $T(\Lambda_R)=\Lambda_R$. Moreover, the right projection of $\Gamma$ is $\Lambda^\vee_R$ and thus, we also require $T(\Lambda^\vee_R)=\Lambda^\vee_R$. Lastly, the $(S,T)$ transformation on an arbitrary element of $\Gamma$ gives
$$
(S,T)(v+\rho(u),u)=(S(v)+S(\rho(u)),T(u))=(S(v)+S(\rho(u))-\rho(T(u))+\rho(T(u)),T(u))
$$
where $v\in\Lambda_L$ and $u\in\Lambda^\vee_R$. For $\Gamma$ to be preserved under this $(S,T)$ transformation, we have
\be\label{permutation nontrivial condition}
S(\rho(u))-\rho(T(u))\in\Lambda_L\;\;\text{for all }u\in\Lambda^\vee_R
\ee
It suffices to check this condition on a basis of $\Lambda_R^\vee$ since $\rho$ is a linear transformation. Now, we want the physical spectrum at the extremum under consideration to stay invariant under this transformation and thus, $p^2_R$ should be invariant, which means that $T$ has to be an orthogonal transformation on $\Lambda_R$ (i.e. it preserves the inner products $u_1.u_2$ for $u_1,u_2\in\Lambda^\vee_R$). Now, the group of orthogonal transformations that preserve a given $ADE$ lattice is a semi-direct product of the Weyl group and the automorphisms of the Dynkin diagram \cite{Dynkin1951Automorphisms}. Automorphisms of the Dynkin diagram are the permutations $\s$ of the simple roots such that the Cartan matrix is preserved, i.e.
$$
C_{\s(i)\s(j)}=C_{ij}\text{  for all }i,j
$$
These Dynkin diagram automorphisms naturally act on the dual basis vectors $\om_i$, defined as
\be\label{dual basis vectors def}
\om_i\cdot\al_j=\d_{ij}
\ee
where $\al_j$ is a simple root. The aforementioned result from \cite{Dynkin1951Automorphisms} implies that any $T$ that satisfies $T(\Lambda_{R})=\Lambda_{R}$ can be written as $T=w\circ d$ where $w$ is a member of the Weyl group, and $d$ is a Dynkin diagram automorphism. Let's take $w$ to be a simple Weyl reflection $w_{\al}$ (corresponding to a simple root $\al$) to write \eqref{permutation nontrivial condition} as
\begin{align}
&S(\rho(u))-\rho(w_{\al}\circ d(u))\in\Lambda_L\nonumber\\
\Rightarrow &S(\rho(u))-\rho(d(u))+(\al\cdot d(u))\rho(\al)\in\Lambda_L\nonumber\\
\label{nontrivial only for dynkin}
\Rightarrow &S(\rho(u))-\rho(d(u))\in\Lambda_L
\end{align}
where we used the fact that $d(u)\in\Lambda^\vee_R$ and $\rho(\al)\in\Lambda_L$. We can replace $w_{\al}$ with an arbitrary product of simple Weyl reflections, and we will still get \eqref{nontrivial only for dynkin}. Therefore, we see that we can restrict $T$ to being Dynkin diagram automorphisms. For the irreducible $ADE$ lattices, these automorphisms are listed in Table \ref{tab: Dynkin automorphisms}. For reducible $ADE$ lattices, the compositions of the automorphisms in Table \ref{tab: Dynkin automorphisms} for every component, and the exchange symmetries of identical components make up the set of Dynkin diagram automorphisms.
\begin{table}[]
    \centering
    \begin{tabular}{|c|c|}
         \hline
         $ADE$ Lattice & Dynkin diagram automorphisms\\\hline
         $A_1$& Identity only\\
         $A_n\;(n>1)$&Chain reversal $(\Z_2)$\\
         $D_4$&Permutation of three outer nodes $(S_3)$\\
         $D_n\;(n> 4)$& Exchange of the two fork endpoints $(\Z_2)$ \\
         $E_6$& Exchange of the two equal arms $(\Z_2)$\\
         $E_7,E_8$&Identity only\\\hline
    \end{tabular}
    \caption{Dynkin diagram automorphisms for the irreducible $ADE$ lattices.}
    \label{tab: Dynkin automorphisms}
\end{table}

Lastly, since $\delta_R=0$, we have
\be
(\delta_L,0)\longmapsto(S(\delta_L),0).
\ee
and since $2\d_{L}\in\Lambda_L$, we have $2S(\delta_{L})\in\Lambda_L$.

Therefore, for a permutation $S$ to be a symmetry of the full $\Gamma$, we need to ensure that $S(\Lambda_L)=\Lambda_L$ and then find at least one Dynkin diagram automorphism $T$ such that \eqref{permutation nontrivial condition} is satisfied for all the dual basis vectors $\om_i$. If any of these conditions fails, then $S$ isn't a symmetry, and we won't remove the apparently redundant $\d$s caused by that permutation. After determining the permutations which are symmetries of the full $\Gamma$ (we will call them acceptable permutations), we consider the orbits of $\d$s, modulo $\Lambda_L$, related by the acceptable permutations and left Weyl reflections, and retain only one representative from a given orbit. We will give a brief description of our algorithm to find $\d$s in subsection \ref{subsection:3.3}.

There is another possible cause of redundant $\d$s, and it is the presence of the Dynkin diagram automorphisms on the left which aren't the permutations of two identical factors in the root system. For example, if $\Lambda_L=2E_8+D_4$ (first entry in Table \ref{tab: rank 4 D4}), then permuting the three outer nodes of $D_4$ Dynkin diagram can be a genuine symmetry of the full $\Gamma$. We haven't pursued checking and removing redundancies under these kinds of possible symmetries in this paper.

Before we discuss toroidal compactifications, let's review how one classifies possible shift vectors in the ten-dimensional, i.e.~$n=0$, case.
\subsection{Ten-dimensional case}\label{subsection: 3.2}
For $n=0$ there are two possible even self-dual lattices of signature $(16,0)$.  Let us start with $\G=\G_{E_8}\oplus\G_{E_8}$.  We want to classify nonzero vectors $\d\in\hlf\G$ with $\d^2\in\Z$, up to equivalence under lattice shifts and Weyl reflections in each $E_8$ factor.  One can do this by brute force as follows.  First pick a basis $\{\al_1,\cdots,\al_{16}\}$ for $\G$.  Consider the set of all $2^{16}$ vectors obtained as $\sum_{i=1}^{16}c_i\al_i$ where $c_i\in\{0,\hlf\}$.  Finally, determine which sets are equivalent to each other under lattice shifts and Weyl reflections, which is time-consuming but tractable.  A far better algorithm \cite{kac1969automorphisms} is provided by the observation of Kac that for the root lattice of an $ADE$ algebra of rank $N$, a fundamental domain for the equivalence under lattice shifts and Weyl reflections is provided by the conditions
\be
x_i:=\d\cdot\al_i\ge 0,\quad i=1,\cdots,N,\qquad\d\cdot\al_h\le 1,
\ee
where $\al_i$ are a basis of simple roots and $\al_h$ is the highest root.  The $x_i$ can be thought of as the coefficients of $\d$ expanded in the dual basis $\om_i$, and the final condition can be expressed as a linear inequality in the $x_i$.  For instance, for $E_8$ the final inequality reads
\be
3x_1+6x_2+5x_3+4x_4+3x_5+2x_6+4x_7+2x_8\le 1.
\ee
Since $\G_{E_8}$ is self-dual, if we are looking for a vector $\d_1\in\hlf\G_{E_8}$ (so the full $\d$ will be of the form $(\d_1;\d_2)$) we simply require that all the $x_i$ are in $\hlf\Z$.  The conditions reduce the possibilities to only three, and using the explicit expressions for the simple roots and their duals in Table \ref{tab:ADE simple and dual basis vectors} in Appendix \ref{app: ADE lattices}, we have
\begin{eqnarray}
x=(x_1,\cdots,x_8)=(0^8)\quad\Rightarrow\quad & \d_1=(0^8),\\
x=\lp 0^7,\hlf\rp\quad\Rightarrow\quad &\d_1'=(0^7,1),\\
x=\lp 0^5,\hlf,0^2\rp\quad\Rightarrow\quad & \d_1''=(0^6,-\hlf,\hlf).
\end{eqnarray}
For each case, we can compute how the $\mathfrak{e}_8$ algebra breaks, how many vectors there are with length squared less than or equal to two in $\G_{E_8}$ and $\G_{E_8}+\d_1$, how those vectors organize into representations of the surviving algebra, and whether the vectors are even or odd under the action of $\d$ (i.e.~whether $\d\cdot v\in\Z$, so $v$ is even, or $\d\cdot v\in\Z+\hlf$, meaning $v$ is odd).  For the three cases above,
\begin{itemize}
    \item $\d_1=(0^8)$, algebra $\mathfrak{e}_8$, $\G_{E_8}=\G_{E_8}+\d_1$ contains a singlet of length squared zero and 240 roots of length squared two, all even under $\d_1$ of course.  We'll write these as\footnote{We use bold-faced numbers to indicate honest representations of the reduced algebra.  For the roots themselves, they don't transform as a full representation since they need to be combined with the Cartan elements to give the adjoint representation. For most of the rest of the paper, for instance in Table~\ref{tab: rank 1 A1}, we won't bother to specify the representations of states.} $\mathbf{1}_0^+$ and $240_2^+$.
    \item $\d_1'=(0^7,1)$, algebra $\mathfrak{d}_8$, $\G_{E_8}$ contains an even singlet of length zero, 112 even vectors of length squared two (assembling to give the adjoint of $D_8$) and 128 odd vectors of length squared two, transforming as a spinor of $D_8$.  We will write this as $\G_{E_8}\rightarrow \mathbf{1}_0^+, 112_2^+, \mathbf{128}_2^-,\cdots$.  The shifted lattice $\G_{E_8}+\d_1'$ contains 16 even vectors of length squared one and 128 odd vectors of length-squared two, $\G_{E_8}+\d_1'\rightarrow \mathbf{16}_1^+,\mathbf{128}_2^-,\cdots$.
    \item $\d_1''=(0^6,-\hlf,\hlf)$, algebra $\mathfrak{e}_7\oplus\mathfrak{a}_1$, and $\G_{E_8}\rightarrow(\mathbf{1},\mathbf{1})_0^+,(126,\mathbf{1})_2^+,(\mathbf{1},2)_2^+,(\mathbf{56},\mathbf{2})_2^-,\cdots$, and $\G_{E_8}+\d_1''\rightarrow(\mathbf{1},\mathbf{2})_{1/2}^-,(\mathbf{56},\mathbf{1})_{3/2}^+,\cdots$.
\end{itemize}

Combining the two $E_8$ factors, imposing $\d^2\in\Z$, and fixing also the exchange symmetry between the two factors, we find the following possible nonzero shift vectors $\d$.  In each case, we examine the $\G+\d$ spectrum and indicate any tachyons.
\begin{enumerate}
    \item $\d=(0^8;0^7,1)$, $\d^2=1$.  Algebra $\mathfrak{e}_8\oplus\mathfrak{d}_8$.
    \be
    \G\rightarrow(\mathbf{1},\mathbf{1})_0^+,(240,\mathbf{1})_2^+,(\mathbf{1},112)_2^+,(\mathbf{1},\mathbf{128})_2^-,\cdots,\qquad\G+\d\rightarrow(\mathbf{1},\mathbf{16})_1^+,(\mathbf{1},\mathbf{128})_2^-,\cdots.
    \ee
    Tachyons with $m^2=-2$ transforming in $(\mathbf{1},\mathbf{16})$.
    \item $\d=(0^7,1;0^7,1)$, $\d^2=2$.  Algebra $\mathfrak{d}_8\oplus\mathfrak{d}_8$.
    \be
    \G\rightarrow(\mathbf{1},\mathbf{1})_0^+,(112,\mathbf{1})_2^+,(\mathbf{1},112)_2^+,(\mathbf{128},\mathbf{1})_2^-,(\mathbf{1},\mathbf{128})_2^-,\cdots,\qquad\G+\d\rightarrow(\mathbf{16},\mathbf{16})_2^+,\cdots.
    \ee
    No tachyons present.
    \item $\d=(0^6,-\hlf,\hlf;0^6,-\hlf,\hlf)$, $\d^2=1$.  Algebra $\mathfrak{e}_7\oplus\mathfrak{e}_7\oplus\mathfrak{a}_1\oplus\mathfrak{a}_1$.
    \begin{align}
    \G\rightarrow&(\mathbf{1},\mathbf{1},\mathbf{1},\mathbf{1})_0^+,(126,\mathbf{1},\mathbf{1},\mathbf{1})_2^+,(\mathbf{1},126,\mathbf{1},\mathbf{1})_2^+,(\mathbf{1},\mathbf{1},2,\mathbf{1})_2^+,(\mathbf{1},\mathbf{1},\mathbf{1},2)_2^+,\nonumber\\
    \qquad&(\mathbf{56},\mathbf{1},\mathbf{2},\mathbf{1})_2^-,(\mathbf{1},\mathbf{56},\mathbf{1},\mathbf{2})_2^-,\cdots,\nonumber\\
    \G+\d\rightarrow&(\mathbf{1},\mathbf{1},\mathbf{2},\mathbf{2})_1^+,(\mathbf{56},\mathbf{1},\mathbf{1},\mathbf{2})_2^-,(\mathbf{1},\mathbf{56},\mathbf{2},\mathbf{1})_2^-,\cdots.
    \end{align}
    Tachyons with $m^2=-2$ transforming in $(\mathbf{1},\mathbf{1},\mathbf{2},\mathbf{2})$.
\end{enumerate}

The case $\G=\G_{D_{16}}+\langle(\hlf^{16})\rangle$ is a bit more complicated because of the shift vector.  Taking a standard basis of simple roots, the conditions on the coefficients of $\d$ are $x_i\ge 0$ and
\be
x_1+x_2+2\sum_{i=3}^{15}x_i+x_{16}\le 1.
\ee
Furthermore, since $2\d\in\G$ which is integral, it is still true that we require $x_i\in\hlf\Z$.  What changes is that some choices of $x_i$ are related by adding the shift vector $(\hlf^{16})$, and for each choice of $x_i$ which satisfies the constraints, we must ask whether $2\d\in\G$.  Before these last two steps, the possible solutions are to set any one of the $x_i$ to one half with the others being zero, to set one of $x_1$, $x_2$, $x_{16}$ to one with all other $x_i=0$, or to set two of $x_1$, $x_2$, $x_{16}$ to one half with all other $x_i=0$ (twenty-two solutions in total).  Restricting to cases where $2\d\in\G$ and $\d^2\in\Z$, and implementing equivalence under the extra shift, we are left with four nonzero possibilities.
\begin{enumerate}
\item $\d=(1,0^{15})$, $\d^2=1$.  Algebra $\mathfrak{d}_{16}$.
\be
\G\rightarrow\mathbf{1}_0^+,480_2^+,\cdots,\qquad\G+\d\rightarrow\mathbf{32}_1^+,\cdots.
\ee
Tachyons with $m^2=-2$ transforming in $\mathbf{32}$.
\item $\d=(\hlf^4,0^{12})$, $\d^2=1$.  Algebra $\mathfrak{d}_{12}\oplus\mathfrak{d}_4$.
\be
\G\rightarrow(\mathbf{1},\mathbf{1})_0^+,(264,\mathbf{1})_2^+,(\mathbf{1},24)_2^+,(\mathbf{24},\mathbf{8})_2^-,\cdots,\qquad\G+\d\rightarrow(\mathbf{1},\mathbf{8})_1^+,(\mathbf{24},\mathbf{8})_2^-,\cdots.
\ee
Tachyons with $m^2=-2$ transforming in $(\mathbf{1},\mathbf{8})$.
\item $\d=(\hlf^8,0^8)$, $\d^2=2$.  Algebra $\mathfrak{d}_8\oplus\mathfrak{d}_8$.
\be
\G\rightarrow(\mathbf{1},\mathbf{1})_0^+,(112,\mathbf{1})_2^+,(\mathbf{1},112)_2^+,(\mathbf{16},\mathbf{16})_2^-,\cdots,\qquad\G+\d\rightarrow(\mathbf{128},\mathbf{1})_2^+,(\mathbf{1},\mathbf{128})_2^+,\cdots.
\ee
No tachyons present.
\item $\d=(\frac{1}{4}^{16})$, $\d^2=1$.  Algebra $\mathfrak{a}_{15}\oplus\mathfrak{u}_1$ (we will only indicate representations under the nonabelian $\mathfrak{a}_{15}$ factor).
\be
\G\rightarrow\mathbf{1}_0^+,240_2^+,\mathbf{120}_2^-,\ov{\mathbf{120}}_2^-,\cdots,\qquad\G+\d\rightarrow 2\times\mathbf{1}_1^+,\mathbf{120}_2^-,\ov{\mathbf{120}}_2^-,\cdots.
\ee
Tachyons with $m^2=-2$ transforming in $2\times\mathbf{1}$.
\end{enumerate}
It is probably not a surprise to the reader that the two tachyon-free $\mathfrak{d}_8\oplus\mathfrak{d}_8$ constructions are equivalent (though as presented we must permute spacetime spinor chiralities to implement the equivalence).  The other five theories round out the set of six non-supersymmetric ten-dimensional heterotic theories which can be obtained by this construction (one other non-supersymmetric construction exists, but cannot be obtained with this construction).

\subsection{$S^1$ compactifications (Rank one)}\label{subsection:3.3}

Now, we find constructions for the rank one case where we perform a supersymmetric breaking orbifold of the heterotic string on $S^1$. Since we are taking $\Lambda_R$ as an $ADE$ lattice, the only choice for a rank one $\Lambda_R$ is $A_1$. The algorithms mentioned before give four different $\Lambda_L$s whose quadratic form is isomorphic to this $\Lambda_R$ (also see \cite{chenevier2026unimodularlatticesrank29} for similar results). These $\Lambda_L$s are given in Table \ref{tab: rank 1 A1}. Next we choose a linear map $\rho:\R\rr\R^{17}$ (which can be specified by providing the image of the generator of $\Lambda_R$), and a vector $\d=(\d_L\vert0)$ with $\d_L\in\hlf\La_L$, that satisfy all the required conditions. For this rank one case, we can either use Kac's algorithm or an algorithm along the lines of the brute force algorithm described in subsection \ref{subsection: 3.2}. We used an implementation of this second algorithm in Mathematica for the higher ranks and thus, let's take a moment to describe this algorithm.

The inputs for this algorithm include the simple roots for the root system in $\Lambda_L$ (whose rank $r$ may be such that $r<16+n$), the basis for the extra lattice (if required, with $16+n-r$ basis vectors), the set of shift vectors, the map $\rho$, and the basis vectors for $\Lambda_R$. We input the map $\rho$ by providing the maps of the dual basis vectors $\om_i$ for $\Lambda_R$ (i.e., the basis vectors for $\Lambda_R^\vee$). We refer to these maps as $\{\rho_1,\cdots,\rho_n\}$. The convention we use for the dual basis vectors for $ADE$ lattices is given in Table \ref{tab:ADE simple and dual basis vectors}. The Mathematica code then constructs a set of basis vectors for $\Lambda_L$, i.e., $B=(b_1,...,b_{16+n})$, where $b_i$ are the basis vectors and $B$ is a matrix with the basis vectors as columns. Then it calculates the Gram matrix $G$ for $\Lambda_L$, i.e., $G=B^TB$, and enumerates all non-trivial $\d_L$ modulo $\Lambda_L$, which are $\d_L=BV$ where $V$ is a vector with entries $0$ or $1/2$ (not all entries of $V$ can be zero, as we aren't considering trivial $\d_L$s). Afterwards, it retains only those $\d_L$ such that $\d^2_L\in\Z$. Finally, it constructs the orbits (modulo $\Lambda_L$) of these $\d_L$ under Weyl reflections and the aforementioned acceptable permutations, and keeps one representative from each orbit. Remember that we are setting $\d_R=0$ and thus, we don't need to classify the right-moving parts of $\d$. The full $\d$ is then $(\d_L\vert0)$. The results of this algorithm agree with the results coming from Kac's algorithm for the $S^1$ case.

Given a choice for $\rho$ and $\d$, the roots of the gauge algebra are given by the vectors $v$ of length-squared two in $\Lambda_L$ which additionally satisfy $\d\cdot v\in\Z$ (we call these $\d$-even vectors). The results for $S^1$ are tabulated in Table \ref{tab: rank 1 A1} in Appendix \ref{app: enhancements}.

\subsection{Higher ranks}

For $T^2$ compactifications, there are two possibilities for $\Lambda_R$, i.e., $A_2$ and $2A_1$. It turns out that for $\Lambda_R=A_2$, we have six different $\Lambda_L$s, and for $\Lambda_R=2A_1$ we have thirteen different $\Lambda_L$s. The root systems for these $\Lambda_L$s, with the number of enhancements, are given in Tables \ref{tab: rank 2 A2} and \ref{tab: rank 2 2A1}. The number of enhancements with different ranks is summarized in Table \ref{tab: T^2 enhancement ranks}. The set of $\d$ for $\Lambda_R$ of rank two and higher is calculated using the algorithm mentioned in the previous subsection.
\begin{table}
    \centering
    \begin{tabular}{|c|c|c|c|c|c|c|c|}
         \hline
         \multicolumn{2}{|c|}{Ranks} & 18 & 17 & 16 & 15 & 14 & Total\\\hline
         \multirow{3}{*}{$A_2$}&Enhancements & 9 & 20 & 19 & 3 & 0 & 51\\
         &Tachyon free& 2 & 0 & 0 & 0 & 0 & 2\\
         &Not knife edges& 2 & 0 & 0 & 0 & 0 & 2\\\hline
         \multirow{3}{*}{$2A_1$}& Enhancements & 118 & 35 & 35 & 13 & 1 & 202\\
         &Tachyon free& 17 & 2 & 0 & 0 & 0 & 19 \\
         &Not knife edges& 2 & 0 & 0 & 0 & 0 & 2 \\\hline
    \end{tabular}
    \caption{Number of enhancements with different ranks in $T^2$ compactifications for $\Lambda_R$ being an $ADE$ lattice.}
    \label{tab: T^2 enhancement ranks}
\end{table}

For $T^3$ compactifications, there are three possibilities for $\Lambda_R$, i.e., $A_3,\;A_2+A_1$, and $3A_1$. It turns out that these possibilities allow $16,27$, and $34$ different $\Lambda_L$, respectively. The root systems for these $\Lambda_L$, with the number of enhancements, are given in Tables \ref{tab: rank 3 A3}, \ref{tab: rank 3 A2+A1}, and \ref{tab: rank 3 3A1} respectively. The number of enhancements with different ranks is summarized in Table \ref{tab: T^3 enhancement ranks}.
\begin{table}
    \centering
    \begin{tabular}{|c|c|c|c|c|c|c|c|c|}
         \hline
         \multicolumn{2}{|c|}{Ranks} & 19 & 18 & 17 & 16 & 15 & 14 & Total\\\hline
         \multirow{3}{*}{$A_3$}& Enhancements & 65& 99 &60  &66  &13  &0  &303 \\
         &Tachyon free&14 &4  &0 &0  &0  &0  &18\\
         &Not knife edges&2 &0  &0 &0  &0  &0  &2\\\hline
         \multirow{3}{*}{$A_2+A_1$}& Enhancements &27 &178  &203 &196 &90 &5  &699 \\
         &Tachyon free&2 &13  &2  &0 &0  &0  &17 \\
         &Not knife edges&2 &13  &2  &0 &0  &0  &17 \\\hline
         \multirow{2}{*}{$3A_1$}&Enhancements & 543 & 111 & 258 & 264 & 151 & 134 & 1461 \\
         &Tachyon free& 64 & 15  & 7 & 6  & 3  & 0  & 95 \\
         &Not knife edges&4 &0  &0 &0  &0  &0  &4 \\\hline
    \end{tabular}
    \caption{Number of enhancements with different ranks in $T^3$ compactifications for $\Lambda_R$ being an $ADE$ lattice.}
    \label{tab: T^3 enhancement ranks}
\end{table}

For $T^4$ compactifications, there are six possible $\Lambda_R$ i.e. $A_4,\;A_3+A_1,\;2A_2,\;A_2+2A_1,\;4A_1$, and $D_4$. It turns out that these $\Lambda_R$ allow $27,\;79,\;51,\;125,\;123$, and $18$ different $\Lambda_L$ respectively. Since the computational load increases significantly with the number of $\Lambda_L$, we provide results for $\Lambda_R=A_4,\;2A_2$, and $D_4$ only, as they have the least number of $\Lambda_L$s. The root systems for the $\Lambda_L$ that appear for these $\Lambda_R$, along with the number of enhancements, are given in Tables \ref{tab: rank 4 A4}, \ref{tab: rank 4 2A2}, and \ref{tab: rank 4 D4} respectively, and the number of enhancements with different ranks are summarized in Table \ref{tab: T^4 enhancement ranks}. One interesting result for the $T^4$ extrema is that there are some extrema which are non-knife-edge, tachyon-free extrema, and they have negative values of the potential in \eqref{potential for Tn}. It is easier to see that these negative potential points are not knife edges because they appear for $\Lambda_R=2A_2$ and $A_4$, which, as we discussed before, can not give knife edges.

A few comments about the format of the ancillary files are in order. The results for $T^n$ with $n=2,3$ and $4$ are provided in the ancillary Excel files for this paper. Each value of $n$ gets one Excel file, and in each of these files, we provide one sheet for each $\Lambda_R$ we study. In each sheet, we list the possible $\d$s; for each $\d$, we provide the corresponding enhanced gauge algebra, the tachyonic $m^2$ that appear in the spectrum (if any, without mentioning the number of such tachyonic states), and the rank of the gauge algebra. We write the non-maximal enhancements (and their ranks) in red. This data is provided for all the root systems possible for the concerned $\Lambda_R$. For each root system, we also provide the basis of the extra lattice if the rank of the root system is less than the maximal rank, and a set of shift vectors which is required to get a root system whose discriminant group is the same as the concerned $\Lambda_R$. We write the extra lattice as the span of the basis vectors for this extra lattice. For example, $E_7+D_{10}$ is a rank 17 root lattice that appears in $T^2$ compactifications, and it requires a one-dimensional lattice, which turns out to be $\sqrt{6}\Z$. We will write the extra lattice as in the Excel file as $\{\{\sqrt{6}\}\}$ for brevity. For a higher-dimensional example, consider the root system $E_6+A_{11}$ that appears for $\Lambda_R=A_2+A_1$. Its extra two-dimensional lattice is provided as $\{\{2,0\},\{0,\sqrt{6}\}\}$ (i.e. the basis vectors for the extra lattice in the extra directions). The map $\rho$ for the lattice $\Lambda_L$ in which the concerned root system sits is also provided in these files. A fourth Excel file contains only the tachyon-free points, with one sheet for each value of $n$ and arranged with respect to $\Lambda_R$. In this file we add a column that records ``No" for all those points which aren't knife edges (i.e. the spectrum at any of these extrema doesn't have a state with $p^2_R=1$ and $p^2_L=2$) and provide the values of the potential \eqref{potential for Tn}, smallest eigenvalue of the Hessian \eqref{Hessian def}, and the ratio of the smallest eigenvalue to the potential for these non-knife edge, tachyon-free points.
\begingroup
\footnotesize
\setlength{\tabcolsep}{5pt}
\setlength{\LTcapwidth}{\textwidth}
\begin{table}
    \begin{tabular}{|c|c|c|c|c|c|c|c|c|c|c|c|c|c|}
         \hline
         \multicolumn{2}{|c|}{Ranks} & 20 & 19 & 18 & 17 & 16 & 15 & 14 & 13 & 12 & 11 & 10 & Total\\\hline
         \multirow{3}{*}{$A_4$}&Enhancements &6 &103 &243 &283 &230  &50 &0 & 0 & 0 & 0 & 0 &  915 \\
         &Tachyon free&2 &13 &2  & 0 &0 &0  & 0 & 0 & 0 & 0 & 0 &  17 \\
         &Not knife edges&2 &13 &2  & 0 &0 &0  & 0 & 0 & 0 & 0 & 0 &  17 \\\hline
         \multirow{3}{*}{$2A_2$}& Enhancements & 17 & 62 & 358 & 740 & 1292 & 1006 & 613 & 204 & 60 & 0 & 1 & 4353  \\
         &Tachyon free& 4 & 7 & 39 & 2  & 13 & 0 & 0 & 0 & 0 & 0 & 0 & 65  \\
         &Not knife edges& 4 & 7 & 39 & 2  & 13 & 0 & 0 & 0 & 0 & 0 & 0 & 65  \\\hline
         \multirow{3}{*}{$D_4$}&Enhancements &243 &75 & 89& 75 &121 &7 & 0 & 0 & 0 & 0 & 0 &  610 \\
         &Tachyon free& 169 & 55 & 63 & 55 & 87 & 4 & 0 & 0 & 0 & 0 & 0 &  433 \\
         &Not knife edges&2 &0 &0  &0  &0 &0  &0 &0 & 0 & 0 & 0 &  2 \\\hline
    \end{tabular}
    \caption{Number of enhancements with different ranks in $T^4$ compactifications for $\Lambda_R$ being $A_4,\;2A_2,\;$ or $D_4$.}
    \label{tab: T^4 enhancement ranks}
\end{table}
\endgroup

\section{Hessians and BF bounds}\label{sec 4: Hessian}
In this section, we will calculate the Hessians of the tachyon-free points which are non-knife edges. We will use these results to test for the BF bound violations of the $AdS_3\times S^3\times T^4$ backgrounds studied in \cite{Robbins:2025wlm}.
\subsection{The set up}
We start by considering $T^n$ compactifications. The $(10-n)$-dimensional effective action in the string frame for $SO(16)\times SO(16)$ heterotic theory on $T^n$ is
\be\label{6d eff action string}
S_{10-n}=\frac{1}{2\k^{2}_{10-n}}\int d^{10-n}x\sqrt{-\widehat{g}}\ls e^{-2\phi} \lp \widehat{R}+4(\p\phi)^{\widehat{2}}-\frac{1}{12}|H_{3}|^{\widehat{2}}-\frac{1}{4}|F_{2}|^{\widehat{2}}-\hlf (\p\s)^{\widehat{2}}\rp-\Lambda-\hlf\mu_{\al\beta}\s^{\al}\s^{\beta}\rs
\ee
where $\phi$ is the $10-n$ dimensional dilaton fluctuation, $\s^{\al}$ are the fluctuations of the $T^{n}$ moduli around an extremum point in the $T^{n}$ moduli space (let's call this point $\s_{*}$). The hats refer to the fact that we are in the string frame. Here, $\k^2_{10-n}$ is given as
\be
\k^2_{10-n}=\frac{\k^2_{10}}{v\al'^{n/2}}=\frac{(2\pi)^7\al'^{4-n/2}g^2_{\circ}}{2v}
\ee
where the convention for $\k^2_{10}$ is consistent with \cite{Polchinski:1998rr} and $v$ is a parameter which specifies the volume of $T^{n}$ as
\be\label{v par def}
\text{volume}(T^{n})=v\al'^{n/2}
\ee
Moreover, $\Lambda=2\la g^{2}_{\circ}/\al'$ is the one-loop correction to the potential\footnote{This $\Lambda$ should not be confused with the notation for lattices used in this paper.} (with $g_{\circ}$ being the one loop corrected string coupling), calculated at $\s_*$, with $\la$ given as
\be\label{lambda expression}
\la=\frac{(2\pi)^7\al'^{5-n/2}}{2v}V
\ee
where $V$ is given in \eqref{potential for Tn}. If we go to Einstein frame using the transformation $\widehat{g}\rr e^{4\phi/(8-n)}g$, \eqref{6d eff action string} becomes
\begin{align}
S_{10-n}=\frac{1}{2\k^{2}_{10-n}}\int d^{10-n}x\sqrt{-g}\ls  R-\frac{4}{8-n}(\p\phi)^{2}-\frac{e^{-8\phi/(8-n)}}{12}|H_{3}|^{2}-\frac{e^{-4\phi/(8-n)}}{4}|F_{2}|^{2}\right.\nonumber\\
\label{6d eff action}
\left.-\hlf (\p\s)^{2}-e^{2(10-n)\phi/(8-n)}\Lambda-\hlf e^{2(10-n)\phi/(8-n)}\mu_{\al\beta}\s^{\al}\s^{\beta}\rs
\end{align}
The $\m_{\al\beta}\s^{\al}\s^{\beta}$ term in \eqref{6d eff action} comes from the dependence of the one-loop correction to the potential on the torus moduli, and the expression for $\m_{\al\beta}$ is
\be\label{mu expression}
\m_{\al\beta}=\frac{(2\pi)^7g^2_{\circ}}{v}\al'^{4-n/2}\frac{\p^2V}{\p\s^\al\p\s^\beta}=\frac{(2\pi)^7g^2_{\circ}}{v}\al'^{4-n/2}H_{\al\beta}
\ee
where we defined the Hessian as
\be\label{Hessian def}
H_{\al\beta}=\frac{\p^2V}{\p\s^\al\p\s^\beta}=-(4\pi^2\al')^{-\frac{10-n}{2}}\int_{\mathcal{F}}\frac{d^{2}\tau}{2\tau^{2}_{2}}\left.\frac{\p^2 Z(\s,\tau)}{\p\s^{\al}\p\s^{\beta}}\right|_{\s_{*}}
\ee
$\mu_{\al\beta}$ is the mass matrix for the moduli $\s^{\al}$, and the eigenvalues of this mass matrix give the values of squared masses which should be above the BF bound for perturbative stability. Notice that this mass matrix is defined with the moduli $\s^{\al}$ that have the canonical kinetic term in the action \eqref{6d eff action}. We will refer to them as canonical moduli for convenience.

Now, suppose that we have torus moduli in some other basis (let's call these moduli $\widetilde{\s}^{\al}$) which give rise to the following terms in the action
\be\label{sigma tilde action}
S_{10-n}\supset \frac{1}{2\k^{2}_{10-n}}\int d^{10-n}x\sqrt{-g}\lp-\hlf \g_{\al\beta}\p_{\m}\widetilde{\s}^{\al}\p^{\m}\widetilde{\s}^{\beta}-\hlf e^{2(10-n)\phi/(8-n)}\widetilde{\mu}_{\al\beta}\widetilde{\s}^{\al}\widetilde{\s}^{\beta}\rp
\ee
If the relation of these moduli with the canonical moduli is $\widetilde{\s}^{\al}=J^{\al}_{\;\;\beta}\sigma^{\beta}$ for some invertible matrix $J$, then \eqref{sigma tilde action} becomes
\be\label{sigma tilde action with J}
S_{10-n}\supset \frac{1}{2\k^{2}_{10-n}}\int d^{10-n}x\sqrt{-g}\lp-\hlf (J^{T}\g J)_{\al\beta}\p_{\m}\s^{\al}\p^{\m}\s^{\beta}-\hlf(J^{T}\widetilde{\mu} J)_{\al\beta}e^{2(10-n)\phi/(8-n)}\s^{\al}\s^{\beta}\rp
\ee
Comparing \eqref{sigma tilde action with J} with the corresponding terms in \eqref{6d eff action}, we get
\be
J^{T}\g J=\mathbb{I}\efill\text{and}\efill J^{T}\widetilde{\m} J=\m
\ee
which gives us the following relations
\be
\g=L^{T}L\efill\m=(L^{-1})^{T}\widetilde{\m}L^{-1}
\ee
where $L$ is an upper triangular matrix. The decomposition of $\g$ given above is called Cholesky's decomposition. A similar relationship would hold between the matrices $H$ and $\wt{H}$ (where $\wt{H}$ is the same as \eqref{Hessian def}, but $\s^\al$ replaced by $\wt{\s}^\al$).

Notice that to check the BF bounds, one needs to find $\m$ (i.e. the mass matrix with respect to the canonical moduli) and to get $\m$, one needs to find $\g$. However, if one is only interested in finding whether the lowest eigenvalue of $H_{\al\beta}$ is negative or not, one doesn't need to find $\g$ and $H_{\al\beta}$ because we can just calculate the lowest eigenvalue of $\wt{H}_{\al\beta}$ with respect to some moduli $\wt{\s}^\al$, and use Sylvester's law of inertia\footnote{Sylvester law of inertia says that if $H$ is a symmetric matrix and $L$ is invertible, then $L^TH L$ and $H$ will have the same number of positive, negative, and zero eigenvalues.} from matrix algebra to deduce that if the lowest eigenvalue of $\wt{H}$ is negative, then so is the case for $H$.

We will use the parametrization of the torus moduli used in subsection \ref{subsection: One-loop potential} i.e., via the $(16+n)\times n$ matrix $M_{ia}$. The second derivatives of $\s_V$ in this parametrization are given as
\be
\left.\frac{\p^2\s_V}{\p M_{ia}\p M_{jb}}\right|_{M=0}=\left.4\pi\tau_2\sum_{\substack{(p_L|p_R)\in\Gamma\\\d.(p_L|p_R)\in\Z}}(4\pi\tau_2p_{Li}p_{Lj}p_{Ra}p_{Rb}-p_{Li}p_{Lj}\d_{ab}-p_{Ra}p_{Rb}\d_{ij})q^{\hlf P^2_L}\bq^{\hlf p^2_R}\right|_{M=0}
\ee
with similar expressions being true for $\s_S,\s_C$ and $\s_O$. The $M_{ia}$ parametrization of the torus moduli gives quite compact expressions for the second derivatives. However, we still need to find $\gamma$ for these moduli. We will show that the tree-level scalar kinetic term, expanded about the reference point $M=0$ and expressed in the Einstein-frame normalization, is
\be\label{eq:result}
L_{\text{kin}}=-\partial_\mu M_{ia}\,\partial^\mu M_{ia}
 =-\hlf\partial_\mu\sigma_{ia}\,\partial^\mu\sigma_{ia},
\ee
where $\s_{ia}$ are some canonical moduli with $\s_{ia}=\sqrt{2}M_{ia}$, and the repeated indices in \eqref{eq:result} are summed over. Thus, we will show that the $\gamma$ metric for the $M_{ia}$ moduli at the reference point $M=0$ is
\begin{equation}\label{eq:cvalue}
 \gamma_{ia,jb}(M=0)=2\,\delta_{ij}\delta_{ab}.
\end{equation}
These statements hold for the quadratic fluctuation action. They do not assert that a constant linear field redefinition makes the kinetic terms canonical throughout the curved Narain moduli space. The local continuous Narain scalar manifold is the symmetric space
\begin{equation}
 \frac{O(16+n,n)}{O(16+n)\times O(n)}.
\end{equation}
In an orthonormal left/right-moving frame, let
\be
\eta=\begin{pmatrix}I_{16+n}&0\\0&-I_n\end{pmatrix},\efill
\mathcal{B}(M)=
\begin{pmatrix}
A(M)&M\\[2pt]
M^T&D(M)
\end{pmatrix},
\ee
where
\be\label{eq:boost}
A(M)=\sqrt{I_{16+n}+MM^T}\efill D(M)=\sqrt{I_n+M^TM}.
\ee
The deformed and undeformed momenta are then related as
\be\label{deformed momenta matrix}
P'=\mathcal{B}(M)P\Rightarrow \begin{pmatrix}
    p'_L\\
    p'_R
\end{pmatrix}=\mathcal{B}(M)\begin{pmatrix}
    p_L\\
    p_R
\end{pmatrix}
\ee
where $P$ is the vector containing $p_L$ and $p_R$. Notice that $\mathcal{B}(0)=I_{16+2n}$. Equation \eqref{deformed momenta matrix} is precisely the $O(16+n,n)$ transformation used in Eq. \eqref{momentum deformtation}. If we want the Lorentzian norm $p^2_L-p^2_R$ to be invariant under the deformation performed by $M$, then we have
\be\label{B with eta}
P'^{T}\eta P'=P^T\eta P\Rightarrow \mathcal{B}(M)^T\eta\mathcal{B}(M)=\eta.
\ee
Using the definition of $\mathcal{B}(M)$ in \eqref{B with eta}, we get the following identities
\be
 A^2-MM^T=I_{16+n},\qquad D^2-M^TM=I_n,\qquad AM=MD
\ee
Although the Lorentzian squared norm $p^2_L-p^2_R$ is preserved under the deformation \eqref{deformed momenta matrix}, the positive definite Euclidean squared norm $p^2_L+p^2_R$ isn't. The deformation of this norm can be written as 
$$
p'^2_L+p'^2_R=P'^TP=P^T\;\mathcal{B}(M)^T\mathcal{B}(M)P=P^T\;\mathcal{H}(M)P
$$
where we defined the symmetric generalized metric
\begin{equation}\label{eq:Hdef}
 \mathcal{H}(M)=\mathcal{B}(M)^T\mathcal{B}(M).
\end{equation}
It satisfies $\mathcal{H}(0)=I_{16+2n}$, is positive definite, and obeys $\mathcal{H}\eta\mathcal{H}=\eta$. Now, for small $M$, we have
\begin{equation}
 A(M)=I_{16+n}+\frac12MM^T+O(M^4),\qquad
 D(M)=I_n+\frac12M^TM+O(M^4).
\end{equation}
We introduce the following symmetric off-diagonal matrix
\begin{equation}\label{eq:Xdef}
 X(M)=\begin{pmatrix}0&M\\M^T&0\end{pmatrix}.
\end{equation}
to get
\begin{equation}\label{eq:expansion}
 \mathcal{B}(M)=I_{16+2n}+X(M)+O(M^2),\qquad
 \mathcal{H}(M)=I_{16+2n}+2X(M)+O(M^2).
\end{equation}
Accordingly, to the order relevant for the kinetic term quadratic in the fluctuations, we have
\begin{equation}\label{eq:derivatives}
 \partial_\mu\mathcal{H}=2\partial_\mu X+O(M\partial M),
 \qquad
 \partial_\mu\mathcal{H}^{-1}=-2\partial_\mu X+O(M\partial M).
\end{equation}
With the standard normalization of the tree-level toroidal scalar action, the generalized-metric term inside the lower-dimensional Einstein-frame Lagrangian is \cite{Maharana_1993}
\begin{equation}\label{eq:cosetkin}
 L_{\mathrm{kin}}=\frac18
 \Tr\!\left(\partial_\mu\mathcal{H}^{-1}\,\partial^\mu\mathcal{H}\right).
\end{equation}
Inserting \eqref{eq:derivatives} gives the quadratic part in $M$ as
\be\label{eq:kin1}
L_{\text{kin}}^{(2)}
=\frac18\Tr \lp(-2\partial_\mu X)(2\partial^\mu X)\rp =-\frac12\Tr \left(\partial_\mu X\,\partial^\mu X\right)=-\;\partial_\mu M_{ia}\,\partial^\mu M_{ia}.
\ee
Comparing this with the definition
\be
L_{\text{kin}}^{(2)}=-\frac12\gamma_{ia,jb}(M=0)\partial_\mu M_{ia}\,\partial^\mu M_{jb}
\ee
we obtain
\begin{equation}
 \gamma_{ia,jb}(M=0)=2\delta_{ij}\delta_{ab}.
\end{equation}
Therefore, the canonical Hessian should be a half of the Hessian obtained via the $M_{ia}$ moduli. We present the values of these canonical Hessians for $T^2, T^3$ and $T^4$ points in Tables \ref{tab:nonknife-T2}, \ref{tab:nonknife-T3}, and \ref{tab:nonknife-T4} respectively. Our results for $T^2$ match with \cite{honda2026}.
\subsection{Hessians and BF bounds for $AdS_3\times S^3\times T^4$}
\begin{figure}
    \centering
    \includegraphics[width=0.5\linewidth]{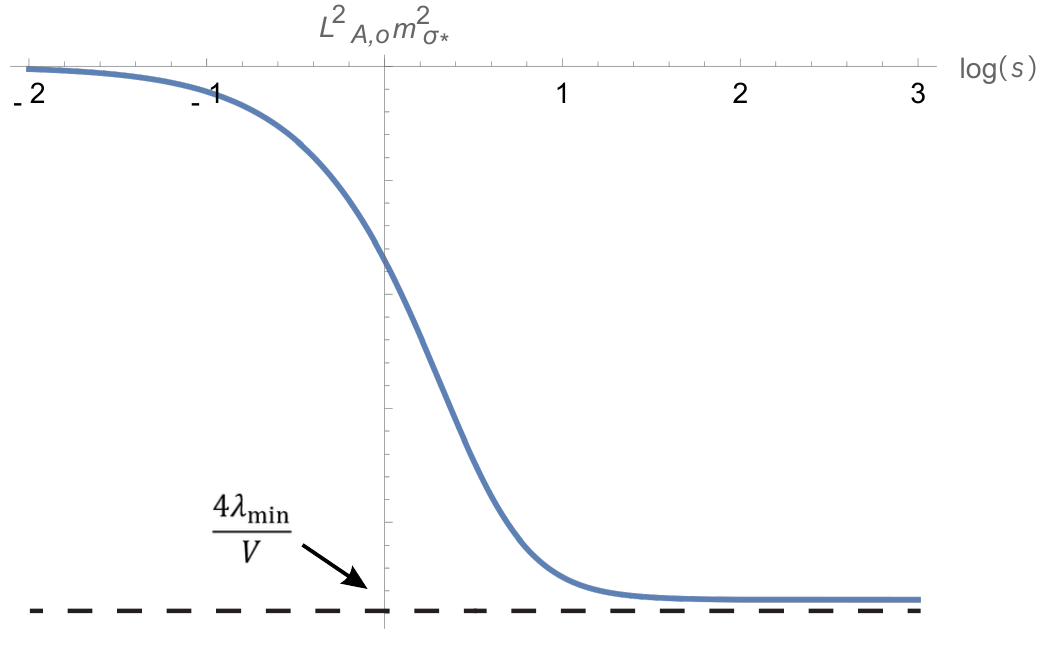}
    \caption{The qualitative shape of the LHS of \eqref{BF bound with s} when $4\lambda_{\text{min}}/V<0$. The horizontal scale is chosen to be the log scale for better scaling. The plot asymptotically approaches $4\lambda_{\text{min}}/V$ for large $s$.}
    \label{fig: qualitative shape}
\end{figure}

Now, we check for the BF bound violations of the $AdS_3\times S^3\times T^4$ backgrounds studied in \cite{Robbins:2025wlm}, and thus, we resort to the $n=4$ case. Using \eqref{mu expression}, it is straightforward to see that the lowest eigenvalue of $\m_{\al\beta}$ (let's call this eigenvalue $m^{2}_{\s_{*}}$) for $n=4$ can be written as
\be\label{m^2_sigma with s}
m^{2}_{\s_{*}}=\la_{\text{min}}\frac{(2\pi)^7 g^{2}_{\circ}}{v}\al'^{4-n/2}\xlongrightarrow{n=4}\la_{\text{min}}\frac{(2\pi)^7 g^{2}_{\circ}\al'^2}{v}
\ee
where $\la_{\text{min}}$ is the lowest eigenvalue of the Hessian\footnote{ This $\la_{\text{min}}$ should not be confused with $\la$ in \eqref{lambda expression}.}. For the $AdS_{3}\times S^{3}\times T^4$ solution studied in \cite{Robbins:2025wlm} the BF bound translates to
\be\label{BF bound def}
L^{2}_{A,\circ}m^{2}_{\s_{*}}\geq -1 
\ee
where $L_{A,\circ}$ is the one loop corrected $AdS$ radius. If any scalar mode violates this bound, then it indicates a perturbative instability in the solution.  This bound would need to be checked for a range of flux pairs $(n_{5},n_{1})$, but we reduce our investigation to a single parameter. To see how to do this, we recall some facts about the aforementioned $AdS_{3}\times S^{3}\times T^4$ solution. This solution has an $H_{3}$ magnetic flux $n_{5}$ on $S^{3}$ and an electric flux $n_{1}$ on $AdS_{3}$. The tree-level solution for string coupling $g_{s}$ and $S^{3}$ radius (denoted as $L$) are
$$
L=\sqrt{\al' n_{5}}\efill\efill g_{s}^{2}=\frac{v |n_{5}|}{(2\pi)^{4}|n_{1}|}
$$
The one-loop corrected string coupling and $S^{3}$ radius are denoted as $L_{\circ}$ and $g_{\circ}$ respectively. These quantities satisfy the following equations
\begin{align}\label{AdS3 eq 1}
2\al'^{2}\lp n^{2}_{5}+\frac{2(2\pi)^{8}g^{4}_{\circ}n^{2}_{1}}{v^{2}}\rp-6L^{4}_{\circ}+\frac{3(2\pi)^{4}\la g^{2}_{\circ}L^{6}_{\circ}}{\al'v}&=0\\ \label{AdS3 eq 2}
2\al'^{2}\lp n^{2}_{5}+\frac{(2\pi)^{8}g^{4}_{\circ}n^{2}_{1}}{v^{2}}\rp-4L^{4}_{\circ}+\frac{(2\pi)^{4}\la g^{2}_{\circ}L^{6}_{\circ}}{\al'v}&=0
\end{align}
and in terms of these quantities, the one loop corrected $AdS$ radius $L_{A,\circ}$ is given as
\be\label{AdS relation}
\frac{1}{L^2_{A,\circ}}=\frac{1}{L^2_{\circ}}-\frac{\Lambda}{4}\Rightarrow L^2_{A,\circ}=\frac{4L^2_{\circ}}{4-L^2_{\circ}\Lambda}
\ee
If we define $s=\frac{\la vn^{2}_{5}}{n_{1}(2\pi)^4}$ and the following dimensionless quantities, which are the ratios of the one-loop corrected quantities with their tree-level counterparts
\be
\xi=\frac{L_{\circ}}{L}\efill \zeta=\frac{g_{\circ}}{g_{s}}
\ee
then it can be shown that the bound in \eqref{BF bound def} can be written as
\be\label{BF bound with s}
L^{2}_{A,\circ}m^{2}_{\s_{*}}\geq -1 \Rightarrow \frac{4\la_{\text{min}}}{V}\frac{s\zeta^2\xi^2}{2-s\zeta^2\xi^2}\geq -1
\ee
where we also used \eqref{lambda expression} to get this result. In the limit of $s$ going to zero, $L_{A,\circ}^2m_{\s_\ast}^2$ also goes to zero and in particular does not violate the BF bound.  Using the large $s$ limits given in \cite{Robbins:2025wlm}, we get the following limit
\be
\lim_{s\rr\infty}\frac{4\la_{\text{min}}}{V}\frac{s\zeta^2\xi^2}{2-s\zeta^2\xi^2}=\frac{4\la_{\text{min}}}{V}
\ee
This limit specifies the value at which the LHS of \eqref{BF bound with s} will saturate at large $s$, which corresponds to small or zero flux on the $AdS_3$ factor, which were called ``intrinsically quantum'' solutions in~\cite{Baykara:2022cwj}. It suffices to check if this limit is smaller than $-1$ to check if there is any BF bound violation. When $4\lambda_\text{min}/V$ is negative, then the qualitative shape of the plots of the LHS of \eqref{BF bound with s} is given in Figure \ref{fig: qualitative shape}. Looking through the entries in Table \ref{tab:nonknife-T4}, we see that all the entries that have a negative $\la_{\text{\text{min}}}/V$ ratio violate the BF bound. On the other hand, a small number of entries have a positive value for the ratio $\la_{\text{\text{min}}}/V$. For these entries, the large $s$ limit (which is the $n_1\rr0$ limit) doesn't exist, since the equations \eqref{AdS3 eq 1} and \eqref{AdS3 eq 2} would lead to a negative $L^2_{\circ}$ for negative $\la$. Therefore, the question regarding the BF bound violations becomes irrelevant for these extrema in the large $s$ limit.

\subsection{Refined de Sitter conjecture}
We can also check the refined de Sitter conjecture \cite{Ooguri:2018wrx, Andriot:2018wzk} for the extrema studied in this paper. In our notation, this conjecture requires the following bound to be satisfied for an extremum with positive value of the potential \eqref{potential for Tn}
\be\label{refined de Sitter}
\left.\frac{m^2_{\s_{*}}}{\Lambda}\right|_{\s_{*}}\leq -c
\ee
where $c$ is an order one constant. Using the expression for $m^{2}_{\s_{*}}$ (given in \eqref{m^2_sigma with s}) and $\Lambda$, \eqref{refined de Sitter} becomes (for any $T^n$)
\be\label{de Sitter bound}
\frac{\la_{\text{min}}}{V} \leq-c
\ee
The values of the LHS of \eqref{de Sitter bound} for $T^n$ with $n=2,3,$ and $4$ are tabulated in Tables \ref{tab:nonknife-T2},\ref{tab:nonknife-T3}, and \ref{tab:nonknife-T4} respectively, and one can see that the refined de Sitter conjecture is satisfied for all the points with positive $V$ if we set $c=0.735$, which is of order one. For $AdS_3\times S^3\times T^4$ backgrounds, since all these points also violate the BF bound, the inequality \eqref{de Sitter bound} is also satisfied by $c=0.25$. Therefore, violating the BF bound implies the validity of the refined de Sitter conjecture in this scenario.
\section{Discussion and Outlook}
In this paper, we have determined the extrema with enhancements of different ranks, when $\Lambda_R$ is an $ADE$ lattice and $\d_R=0$. Among these extrema, there are thousands of non-maximal extrema, and hundreds of those non-maximal extrema are tachyon-free. We have also calculated the Hessians for the tachyon-free points that aren't knife edges, and determined that none of these points is a minimum. Among these extrema, there are some points that have a negative value of the one-loop potential, and all of these points are non-maximal extrema. These points can lead to interesting constructions, and it is a future direction that can be pursued.

It is tempting to extend our analysis to the $n=5$ and $6$ cases, with the $n=6$ case being quite interesting because it leads to four large dimensions. This extension can be done in principle, but becomes computationally demanding, as there can be several hundred $\Lambda_L$s that correspond to a given $\Lambda_R$ (thousands in some cases). In that scenario, one can resort to obtaining all tachyon-free enhancements, or those tachyon-free enhancements which are not knife edges.

Another interesting question is to consider more general constructions of the even self-dual lattice $\Gamma$ where the existence of extrema doesn't hinge on the existence of Weyl reflections, but on some other automorphisms of the lattice (or a sublattice). In addition, the issue of classifying maximal enhancements for $T^n$ with $n>2$ is not addressed, and it would be interesting to see if we require a more general lattice construction to get some points in those classifications.

\section*{Acknowledgments}

The authors would like to thank Bernardo Fraiman, Alexander Maloney, Ivano Basile, and Hector Parra de Freitas for the useful discussions and email exchanges regarding this work. D.R.~and H.S.~would like to gratefully acknowledge support from the Simons Center for Geometry and Physics, Stony Brook University, at which some of the research for this paper was performed. S.M.~and H.S.~would like to thank the organizers of ``Mirror Symmetry, Calabi-Yau Threefolds, and Connections to Physics" at ICERM, Brown University, for their hospitality during part of this work, and for facilitating the start of the collaboration between S.M.~and the other two authors. H.S.~would like to thank the organizers of PMQFTSS 2026 at the University of Massachusetts, Amherst, for their hospitality during part of this work. We used ChatGPT-6 Astra (OpenAI) to convert our Mathematica results to Excel files, and D.R.~used Claude Sonnet 5 (Anthropic) for literature searches and some preliminary calculations of lattice genera. No part of this paper was written by AI, and the authors take full responsibility for the text in this paper.
\appendix
\section{ADE lattices}\label{app: ADE lattices}
\begin{itemize}
    \item The $A_N$ lattices are embedded into the subspace of $\R^{N+1}$ where the sum of all components is zero (i.e.~orthogonal to the vector $(1^{N+1})$).  We can write
    \begin{equation}
        \G_{A_N}=\left\{(n_1,\cdots,n_{N+1})\mid n_i\in\Z,\ \sum_{i=1}^{N+1}n_i=0\right\}.
    \end{equation}
    The dual lattice is
    \begin{equation}
        \Gamma_{A_N}^\vee=\Gamma_{A_N}+\left\langle\lp\frac{1}{N+1}^N,-\frac{N}{N+1}\rp\right\rangle,
    \end{equation}
    and the index group is $\G_{A_N}^\vee/\G_{A_N}\cong\Z_{N+1}$. The notation $\langle\{\cdots\}\rangle$ refers to span of the vectors in the set $\{\cdots\}$. If we want to refer to the span of a single vector, we won't use the set brackets. For example, the span of a vector $v$ would be written as $\langle v\rangle$.
    \item The $D_N$ lattices are specified as
    \begin{equation}
        \G_{D_N}=\left\{(n_1,\cdots,n_N)\mid n_i\in\Z,\ \sum_{i=1}^Nn_i\in 2\Z\right\}.
    \end{equation}
    The dual lattice can be written as
    \begin{equation}\label{DN dual}
        \Gamma_{D_N}^\vee=\left\{(n_1,\cdots,n_N)\mid n_i\in\Z\right\}\bigcup\left\{\lp n_1+\hlf,\cdots,n_N+\hlf\rp\mid n_i\in\Z\right\},
    \end{equation}
    though there is a qualitative difference between the cases where $N$ is even or odd.  For $N$ even we can write
    \begin{equation}
        \G_{D_N}^\vee=\G_{D_N}+\left\langle\left\{\lp\hlf\rp^N,(1,0^{N-1})\right\}\right\rangle,\qquad N\text{ even},
    \end{equation}
    with index group $\G_{D_N}^\vee/\G_{D_N}\cong\Z_2^2$, while for $N$ odd we can write
    \begin{equation}
        \G_{D_N}^\vee=\G_{D_N}+\left\langle\lp\hlf\rp^N\right\rangle,\qquad N\text{ odd},
    \end{equation}
    with index group $\G_{D_N}^\vee/\G_{D_N}\cong\Z_4$.
    \item For $E_6$ we will use the convention
    \begin{equation}
        \G_{E_6}=\left\{(n_1,\cdots,n_5,0)\mid n_i\in\Z,\ \sum_{i=1}^5n_i\in 2\Z\right\}+\left\langle\lp\hlf^5,-\frac{\sqrt{3}}{2}\rp\right\rangle.
    \end{equation}
    The dual lattice is
    \begin{equation}
        \G_{E_6}^\vee=\G_{E_6}+\left\langle\lp\hlf^5,\frac{1}{2\sqrt{3}}\rp\right\rangle,
    \end{equation}
    and the index group is $\G_{E_6}^\vee/\G_{E_6}\cong\Z_3$.
    \item For $E_7$ we embed into the $(1^8)$-orthogonal subspace of $\R^8$ as
    \begin{equation}
        \G_{E_7}=\left\{(n_1,\cdots,n_8)\mid n_i\in\Z,\ \sum_{i=1}^8n_i=0\right\}+\left\langle\lp\hlf^4, -\hlf^4\rp\right\rangle.
    \end{equation}
    The dual lattice is
    \begin{equation}
        \G_{E_7}^\vee=\G_{E_7}+\left\langle\lp\frac{1}{4}^6, -\frac{3}{4}^2\rp\right\rangle,
    \end{equation}
    and the index group is $\G_{E_7}^\vee/\G_{E_7}\cong\Z_2$.
    \item Finally, the $E_8$ lattice is given as
    \begin{equation}
        \G_{E_8}=\left\{(n_1,\cdots,n_8)\mid n_i\in\Z,\ \sum_{i=1}^8n_i\in 2\Z\right\}+\left\langle\lp\hlf^8\rp\right\rangle.
    \end{equation}
    This lattice is self-dual, $\G_{E_8}^\vee=\G_{E_8}$, and the index group is trivial.
\end{itemize}
Our conventions for the simple roots $\al_i$ and dual basis vectors $\om_i$ (i.e. $\om_i\cdot\al_j=\d_{ij}$) are given in Table \ref{tab:ADE simple and dual basis vectors}.
\begingroup
\footnotesize
\renewcommand{\arraystretch}{1.6}
\setlength{\tabcolsep}{7pt}
\setlength\LTleft{\fill}
\setlength\LTright{\fill}

\begin{longtable}{|>{\centering\arraybackslash}p{0.15\textwidth}|
                  >{\centering\arraybackslash}p{0.40\textwidth}|
                  >{\centering\arraybackslash}p{0.40\textwidth}|}
\hline
Root lattice & Simple roots $\alpha_i$ & Dual basis vectors $\om_i$ \\
\hline
\endfirsthead




$A_n$
&
$
\alpha_i=e_i-e_{i+1},
\qquad i=1,\ldots,n
$
&
$
\begin{gathered}
\om_k=
\left(
\left(\frac{n+1-k}{n+1}\right)^k,
\left(-\frac{k}{n+1}\right)^{n+1-k}
\right),\\
k=1,\ldots,n
\end{gathered}
$
\\[10pt] \hline


$D_n$
&
$
\begin{gathered}
\alpha_1=(-1,-1,0^{n-2}),\\
\alpha_i=e_{i-1}-e_i,
\qquad i=2,\ldots,n
\end{gathered}
$
&
$
\begin{gathered}
\om_1=\lp-\hlf^n\rp, \om_2=\lp\hlf,-\hlf^{\,n-1}\rp,\\
\om_i=(0^{i-1},(-1)^{n-i+1}),
\qquad i=3,\ldots,n
\end{gathered}
$
\\[10pt] \hline


$E_6$
&
$
\begin{gathered}
\alpha_1=(0^3,1,1,0),\\
\alpha_2=(1,-1,0^4),\\
\alpha_3=(0,1,-1,0^3),\\
\alpha_4=(0^2,1,-1,0^2),\\
\alpha_5=(0^3,1,-1,0),\\
\alpha_6=\left(\hlf^5,-\frac{\sqrt3}{2}\right)
\end{gathered}
$
&
$
\begin{gathered}
\om_1=\left(\hlf^5,\frac{5\sqrt3}{6}\right),\\
\om_2=\left(1,0^4,\frac{1}{\sqrt3}\right),\\
\om_3=\left(1^2,0^3,\frac{2}{\sqrt3}\right),\\
\om_4=\left(1^3,0^2,\sqrt3\right),\\
\om_5=\left(\hlf^4,-\hlf,\frac{\sqrt3}{2}\right),\\
\om_6=\left(0^5,-\frac{2}{\sqrt3}\right)
\end{gathered}
$
\\[10pt] \hline


$E_7$
&
$
\begin{gathered}
\alpha_i=e_{i+1}-e_{i+2},
\qquad i=1,\ldots,6,\\
\alpha_7=\lp\hlf^4,-\hlf^4\rp
\end{gathered}
$
&
$
\begin{gathered}
\om_1=(-1,1,0^6),\\
\om_2=(-2,1^2,0^5),\\
\om_3=(-3,1^3,0^4),\\
\om_4=\left(-\frac94,\frac{3}{4}^4,
-\frac{1}{4}^3\right),\\
\om_5=\left(-\frac32,\hlf^5,-\hlf^2\right),\\
\om_6=\left(-\frac{3}{4},\frac{1}{4}^6,-\frac{3}{4}\right),\\
\om_7=\left(\frac{7}{4},-\frac{1}{4}^7\right)
\end{gathered}
$
\\[10pt] \hline


$E_8$
&
$
\begin{gathered}
\alpha_i=e_i-e_{i+1},
\qquad i=1,\ldots,6,\\
\alpha_7=\lp-1,-1,0^6\rp,\\
\alpha_8=\lp\hlf^8\rp
\end{gathered}
$
&
$
\begin{gathered}
\om_1=\left(\hlf,-\hlf^6,\frac52\right),\\
\om_2=(0^2,-1^5,5),\\
\om_3=(0^3,-1^4,4),\\
\om_4=(0^4,-1^3,3),\\
\om_5=(0^5,-1^2,2),\\
\om_6=(0^6,-1,1),\\
\om_7=\left(-\hlf^7,\frac72\right),\\
\om_8=(0^7,2)
\end{gathered}
$
\\[10pt] \hline
\caption{Our conventions for the simple roots and dual basis vectors for the ADE root lattices. The dual basis is defined by $\om_i\cdot\alpha_j=\delta_{ij}$. In this table, $e_i$ is a vector with $1$ in the $i-$th entry and zeros elsewhere.}
\label{tab:ADE simple and dual basis vectors}
\end{longtable}
\endgroup
\section{Theta functions}\label{app: theta functions}
The theta functions $V_{8},S_{8},O_{8}$ and $C_{8}$, whose anti-holomorphic versions are mentioned in \eqref{one loop partition function} are defined as
\be
V_8(\tau)=\frac{1}{2\eta(\tau)^{4}}\lp \vartheta^{4}\ls\begin{matrix}0\\0\end{matrix}\rs(0,\tau)-\vartheta^{4}\ls\begin{matrix}0\\1/2\end{matrix}\rs(0,\tau)
\rp,
\ee
\be
S_{8}(\tau)=\frac{1}{2\eta(\tau)^{4}}\lp \vartheta^{4}\ls\begin{matrix}1/2\\0\end{matrix}\rs(0,\tau)+\vartheta^{4}\ls\begin{matrix}1/2\\1/2\end{matrix}\rs(0,\tau)
\rp,
\ee
\be
O_{8}(\tau)=\frac{1}{2\eta(\tau)^{4}}\lp \vartheta^{4}\ls\begin{matrix}0\\0\end{matrix}\rs(0,\tau)+\vartheta^{4}\ls\begin{matrix}0\\1/2\end{matrix}\rs(0,\tau)
\rp,
\ee
\be
C_{8}(\tau)=\frac{1}{2\eta(\tau)^{4}}\lp \vartheta^{4}\ls\begin{matrix}1/2\\0\end{matrix}\rs(0,\tau)-\vartheta^{4}\ls\begin{matrix}1/2\\1/2\end{matrix}\rs(0,\tau)
\rp,
\ee
where $\eta$ is the Dedekind eta function
\be
\eta(\tau)=q^{1/24}\prod_{k=1}^{\infty}\lp 1-q^{k}\rp\;\;\;\text{where }\; q=e^{2\pi i \tau},
\ee
and the generalized theta functions are defined as (see \cite{Polchinski:1998rq} for details)
\be
\vartheta\ls\begin{matrix}
    \al\\\beta\end{matrix}\rs(z,\tau)=\sum_{n\in\Z}\exp\lp \pi i (n+\al)^{2}\tau+2\pi i (n+\al)(z+\beta)\rp.
\ee

\section{Lists of $\Lambda_L$s and enhancement}\label{app: enhancements}

\subsection{Rank 1}
\label{appsubsec:Rank17}

For $\Lambda_R$ of rank 1, there is only one choice, and it is $\Lambda_R=A_1$. For this choice, the list of $\Lambda_L$s, the corresponding $\rho$ (in the form of the map of the generator of $A^{\vee}_1$), the list of $\d$s for every $\Lambda_L$, and the enhancements that they produce are tabulated in Table \ref{tab: rank 1 A1} below. For maximal enhancements, we quote the line number from Table 8 of \cite{Fraiman:2023cpa} where the corresponding entry appears.
\begingroup
\renewcommand{\arraystretch}{1.5}
\setlength{\tabcolsep}{4pt}
\setlength{\LTcapwidth}{\textwidth}
\setlength\LTleft{-2 cm}
\setlength\LTright{\fill}

\begin{longtable}{
|>{\centering\arraybackslash}p{0.20\textwidth}|
>{\centering\arraybackslash}p{0.14\textwidth}|
>{\centering\arraybackslash}p{0.20\textwidth}|
>{\centering\arraybackslash}p{0.15\textwidth}|
>{\centering\arraybackslash}p{0.29\textwidth}|
>{\centering\arraybackslash}p{0.09\textwidth}|}

\hline
$\Lambda_L$ & $\rho$ & $\d$ & Enhancement & Tachyonic $m^2$ & Line no. \\
\hline
\endfirsthead

\hline
$\Lambda_L$ & $\rho$ & $\d$ & Enhancement & Tachyonic $m^2$ & Line no. \\
\hline
\endhead

\hline
\endfoot

\endlastfoot


$2E_8+A_1$
&
$\lp0^8;0^8;\hlf,-\hlf\rp$
&
$\lp 0^8;0^6,\hlf^2;\hlf,-\hlf\rp$
&
$E_8+E_7+2A_1$
&
$(1,1,2,2)_{-2},2(1,56,1,1)_{-1}$
&
$106$
\\[8pt]\cline{3-6}

&
&
$\lp0^8;0^7,1;0^2\rp$
&
$E_8+D_8+A_1$
&
$(1,16,1)_{-2},2(1,16,2)_{-1}$
&
$103$
\\[8pt]\cline{3-6}

&
&
$\lp0^6,\hlf^2;0^6,\hlf^2;0^2\rp$
&
$2E_7+3A_1$
&
$(1,1,2,2,1)_{-2},2(1,1,2,2,2)_{-1}$
&
$97$
\\[8pt]\cline{3-6}

&
&
$\lp0^7,1;0^6,\hlf^2;\hlf,-\hlf\rp$
&
$E_7+D_8+2A_1$
&
$2(1,16,2,1)_{-1}$
&
$94$
\\[8pt]\cline{3-6}

&
&
$\lp0^7,1;0^7,1;0^2\rp$
&
$2D_8+A_1$
&
None
&
$41$
\\[8pt]\hline


$
\begin{gathered}
D_{16}+A_1\\
+\Bigl\langle
\Bigl(\hlf^{16};0^2\Bigr)
\Bigr\rangle
\end{gathered}
$
&
$\lp0^{16};\hlf,-\hlf\rp$
&
$\lp \frac{1}{2}^2,0^{14};\frac{1}{2},-\frac{1}{2}\rp$
&
$D_{14}+3A_1$
&
$(1,1,1,2)_{-2},2(28,2,1,1)_{-1}$
&
66
\\[8pt]\cline{3-6}

&
&
$\lp1,0^{15};0^2\rp$
&
$D_{16}+A_1$
&
$(32,1)_{-2},2(32,2)_{-1}$
&
69
\\[8pt]\cline{3-6}

&
&
$\lp\frac{1}{2}^4,0^{12};0^2\rp$
&
$D_{12}+D_4+A_1$
&
$(1,8,1)_{-2},2(1,8,2)_{-1}$
&
60
\\[8pt]\cline{3-6}

&
&
$\lp\frac{1}{2}^6,0^{10};\frac{1}{2},-\frac{1}{2}\rp$
&
$D_{10}+D_6+A_1$
&
$2(1,32,1)_{-1}$
&
56
\\[8pt]\cline{3-6}

&
&
$\lp\frac{1}{2}^8,0^8;0^2\rp$
&
$2D_{8}+A_1$
&
None
&
41
\\[8pt]\cline{3-6}

&
&
$\lp\frac{3}{4},\frac{1}{4}^{14},-\frac{1}{4};
\frac{1}{2},-\frac{1}{2}\rp$
&
{\color{red}$A_{15}+A_1$}
&
$(16+\overline{16},1)_{-1}$
&
-
\\[8pt]\cline{3-6}

&
&
$\lp\frac{1}{4}^{16};0^2\rp$
&
{\color{red}$A_{15}+A_1$}
&
$(1,1)_{-2},2(1,2)_{-1}$
&
-
\\[8pt]\hline


$
\begin{gathered}
E_7+D_{10}\\
+\Bigl\langle
\Bigl(
\frac{1}{4}^{6},-\frac{3}{4}^2;
\hlf^{10}
\Bigr)
\Bigr\rangle
\end{gathered}
$
&
$\lp0^8;\hlf^9,-\hlf\rp$
&
$\lp0^8;1,0^9\rp$
&
$E_7+D_{10}$
&
$(1,20)_{-2},2(56,1)_{-1}$
&
96
\\[8pt]\cline{3-6}

&
&
$\lp0^8;\frac{1}{2}^4,0^6\rp$
&
$E_7+D_6+D_4$
&
$(1,1,8)_{-2},2(1,32,1)_{-1}$
&
91
\\[8pt]\cline{3-6}

&
&
$\lp0^8;\frac{1}{2}^8,0^2\rp$
&
$E_7+D_8+2A_1$
&
$2(1,16,2,1)_{-1}$
&
94
\\[8pt]\cline{3-6}

&
&
$\lp0^6,\frac{1}{2},-\frac{1}{2};
\frac{1}{2}^2,0^8\rp$
&
$D_8+D_6+3A_1$
&
$(1,1,2,2,1)_{-2},(1,12,1,1,2)_{-1}$
&
39
\\[8pt]\cline{3-6}

&
&
$\lp0^6,\frac{1}{2},-\frac{1}{2};
\frac{1}{2}^6,0^4\rp$
&
$2D_6+D_4+A_1$
&
$2(1,1,8,2)_{-1}$
&
30
\\[8pt]\cline{3-6}

&
&
$\lp0^6,\frac{1}{2},-\frac{1}{2};
\frac{1}{2}^{10}\rp$
&
$D_{10}+D_6+A_1$
&
$(1,12,1)_{-2},2(20,1,2)_{-1}$
&
57
\\[8pt]\cline{3-6}

&
&
$\lp0^6,\frac{1}{2},-\frac{1}{2};
\frac{1}{2}^9,-\frac{1}{2}\rp$
&
$D_{10}+D_6+A_1$
&
$2(1,32,1)_{-1}$
&
56
\\[8pt]\cline{3-6}

&
&
$\lp\frac{1}{8}^6,-\frac{3}{8}^2;
\frac{1}{4}^{10}\rp$
&
{\color{red}$E_6+A_9$}
&
$2(1,1)_{-2},2(1,10+\overline{10})_{-1}$
&
-
\\[8pt]\cline{3-6}

&
&
$\lp\frac{1}{8}^7,-\frac{7}{8};
\frac{3}{4},\frac{1}{4}^8,-\frac{1}{4}\rp$
&
{\color{red}$A_7+A_9$}
&
$(8+\overline{8},1)_{-1}$
&
-
\\[8pt]\hline


$
\begin{gathered}
A_{17}\\
+\Bigl\langle
\Bigl(
\frac{1}{3}^{12},-\frac{2}{3}^6
\Bigr)
\Bigr\rangle
\end{gathered}
$
&
$\lp\frac{1}{6}^{15},-\frac{5}{6}^3\rp$
&
$\lp\hlf^2,0^{14},-\hlf^2\rp$
&
{\color{red}$A_{13}+A_3$}
&
$(1,6)_{-2},(14+\overline{14},1)_{-1}$
&
-
\\[8pt]\cline{3-6}

&
&
$\lp\hlf^4,0^{10},-\hlf^4\rp$
&
{\color{red}$A_9+A_7$}
&
$(1,8+\overline{8})_{-1}$
&
-
\\[8pt]\cline{3-6}

&
&
$\lp\hlf^6,0^6,-\hlf^6\rp$
&
{\color{red}$A_{11}+A_5$}
&
$(1,1)_{-2},2(1,20)_{-1}$
&
-
\\[8pt]\cline{3-6}

&
&
$\lp\hlf^8,0^2,-\hlf^8\rp$
&
{\color{red}$A_{15}+A_1$}
&
$(16+\overline{16},2)_{-1}$
&
-
\\[8pt]\hline
\caption{The list of $\Lambda_L$, $\rho$, $\d$, enhancements, and tachyonic spectra, for $\Lambda_R=A_1$. The non-maximal enhancements are highlighted in red color. For the maximal enhancements, we quote the line number of Table 8 in \cite{Fraiman:2023cpa} where they appear.}
\label{tab: rank 1 A1}\\

\end{longtable}
\endgroup

\subsection{Rank 2}
In the tables that follow in the next three subsections, we will provide the root systems for different $\Lambda_R$ for $T^2, T^3$ and $T^4$ compactifications. The basis for the extra lattice mentioned in subsection \eqref{subsection:3.3} and the shift vectors are provided in the ancillary Excel files. The lists of root systems for $\Lambda_R=A_2$ and $2A_1$ are given in Tables \ref{tab: rank 2 A2} and \ref{tab: rank 2 2A1} respectively.
\begingroup
\renewcommand{\arraystretch}{1.6}
\setlength{\tabcolsep}{8pt}
\setlength\LTleft{-1 cm}
\setlength\LTright{\fill}
\begin{longtable}{|>{\centering\arraybackslash}p{0.05\textwidth}|
                  >{\centering\arraybackslash}p{0.78\textwidth}|
                  >{\centering\arraybackslash}p{0.14\textwidth}|}
\hline
\# & Root system & Enhancements \\
\hline
\endfirsthead

\hline
\# & Root system & Enhancements \\
\hline
\endhead

\endfoot


1 &
$
2E_8+A_2
$
&
$5$
\\[8pt] \hline


2 &
$
D_{16}+A_2
$
&
$7$
\\[8pt] \hline


3 &
$
\begin{gathered}
E_7+D_{10}
\end{gathered}
$
&
$14$
\\[8 pt] \hline


4 &
$
3E_6
$
&
$5$
\\[8pt] \hline


5 &
$
D_7+A_{11}
$
&
$12$
\\[8pt] \hline


6 &
$
A_{17}
$
&
$8$
\\[8pt] \hline


&
\textbf{Total enhancements}
&
$\mathbf{51}$
\\[8pt] \hline

\caption{Root systems and number of enhancements for $\Lambda_R=A_2$.}
\label{tab: rank 2 A2}\\
\end{longtable}
\endgroup


\begingroup
\renewcommand{\arraystretch}{1.6}
\setlength{\tabcolsep}{8pt}
\setlength\LTleft{-1 cm}
\setlength\LTright{\fill}
\begin{longtable}{|>{\centering\arraybackslash}p{0.05\textwidth}|
                  >{\centering\arraybackslash}p{0.78\textwidth}|
                  >{\centering\arraybackslash}p{0.14\textwidth}|}
\hline
\# & Root system & Enhancements \\
\hline
\endfirsthead

\hline
\# & Root system & Enhancements \\
\hline
\endhead

\endfoot


1 &
$2E_8+2A_1$
&9 \\[4pt] \hline


2 &
$
A_{17}+A_1
$
&9 \\[8pt] \hline


3 &
$
2A_9
$
&17 \\[8pt] \hline


4 &
$
A_{15}+A_3
$
&14 \\[8pt] \hline


5 &
$
E_6+A_{11}
$
&20 \\[8pt] \hline


6 &
$
2E_7+D_4
$
&17 \\[8pt] \hline


7 &
$
\begin{gathered}
3D_6
\end{gathered}
$
&35 \\[8 pt]\hline


8 &
$
D_{12}+D_6
$
&16 \\[8pt] \hline


9 &
$
E_8+D_{10}
$
&11 \\[4pt] \hline


10 &
$
D_{18}
$
&5 \\[4pt] \hline


11 &
$
\begin{gathered}
2D_8+2A_1
\end{gathered}
$
&18 \\[8 pt]\hline


12 &
$
E_7+D_{10}+A_1
$
&19 \\[8pt] \hline


13 &
$
D_{16}+2A_1
$
&12 \\[8pt]\hline


&
\textbf{Total enhancements}
&\textbf{202}
\\[8pt]\hline
\caption{\mbox{Root systems and number of enhancements for $\Lambda_R=2A_1$}}
\label{tab: rank 2 2A1}
\end{longtable}
\endgroup

\subsection{Rank 3}
The lists of root systems for $\Lambda_R=A_3,\;A_2+A_1$, and $3A_1$ are given in Tables \ref{tab: rank 3 A3}, \ref{tab: rank 3 A2+A1}, and \ref{tab: rank 3 3A1} respectively.
\begingroup
\renewcommand{\arraystretch}{1.6}
\setlength{\tabcolsep}{8pt}
\setlength\LTleft{-1 cm}
\setlength\LTright{\fill}

\begin{longtable}{
|>{\centering\arraybackslash}p{0.05\textwidth}|
>{\centering\arraybackslash}p{0.78\textwidth}|
>{\centering\arraybackslash}p{0.14\textwidth}|}

\hline
\# & Root system & Enhancements \\
\hline
\endfirsthead

\hline
\# & Root system & Enhancements \\
\hline
\endhead

\endfoot


1 &
$
2E_8+A_3
$
&9 \\[4pt] \hline


2 &
$
E_8+D_{11}
$
&10 \\[4pt] \hline


3 &
$
2E_7+D_5
$
&16 \\[8pt] \hline


4 &
$
\begin{gathered}
E_7+D_{10}+A_1
\end{gathered}
$
&24 \\[8 pt] \hline


5 &
$
\begin{gathered}
3E_6
\end{gathered}
$
&11 \\[8pt] \hline


6 &
$
E_6+A_{11}+2A_1
$
&22 \\[8pt] \hline


7 &
$
D_{19}
$
&5 \\[4pt] \hline


8 &
$
D_{16}+A_3
$
&12 \\[8pt] \hline


9 &
$
D_{12}+D_7
$
&16 \\[8pt] \hline


10 &
$
\begin{gathered}
2D_8+A_3
\end{gathered}
$
&20 \\[8 pt] \hline


11 &
$
D_7+A_{11}
$
&25 \\[8pt] \hline


12 &
$
\begin{gathered}
3D_6
\end{gathered}
$
&49 \\[8 pt] \hline


13 &
$
D_5+2A_7
$
&24 \\[8pt] \hline


14 &
$
D_4+A_{15}
$
&18 \\[8pt] \hline


15 &
$
A_{17}+A_1
$
&13 \\[8pt] \hline


16 &
$
2A_9
$
&29 \\[8pt] \hline

&
\textbf{Total enhancements}
&\textbf{303}
\\[8pt] \hline

\caption{Root systems and number of enhancements for $\Lambda_R=A_3$.}
\label{tab: rank 3 A3}\\

\end{longtable}
\endgroup
\begingroup
\renewcommand{\arraystretch}{1.6}
\setlength{\tabcolsep}{8pt}
\setlength{\LTcapwidth}{\textwidth}
\setlength\LTleft{-1 cm}
\setlength\LTright{\fill}
\begin{longtable}{
|>{\centering\arraybackslash}p{0.05\textwidth}|
>{\centering\arraybackslash}p{0.78\textwidth}|
>{\centering\arraybackslash}p{0.14\textwidth}|}

\hline
\# & Root system & Enhancements\\
\hline
\endfirsthead

\hline
\# & Root system & Enhancements\\
\hline
\endhead

\endfoot

1 &
$2E_8+A_2+A_1$
& $11$
\\[8pt]\hline

2 &
$
E_8+D_{10}
$
& $14$
\\[8pt]\hline

3 &
$
\begin{gathered}
2E_7+D_4
\end{gathered}
$
& $31$
\\[8 pt]\hline

4 &
$
E_7+D_{10}+A_2
$
& $19$
\\[8pt]\hline

5 &
$
\begin{gathered}
E_7+D_{10}+A_1
\end{gathered}
$
& $29$
\\[8 pt]\hline

6 &
$
D_{16}+A_2+A_1
$
& $15$
\\[8pt]\hline

7 &
$
\begin{gathered}
D_{12}+D_6
\end{gathered}
$
& $20$
\\[8 pt]\hline

8 &
$
\begin{gathered}
2D_8+2A_1
\end{gathered}
$
& $28$
\\[8 pt]\hline

9 &
$
E_7+A_{11}
$
& $19$
\\[8pt]\hline

10 &
$
D_9+A_9
$
& $23$
\\[8pt]\hline

11 &
$
D_{18}
$
& $6$
\\[8pt]\hline

12 &
$
\begin{gathered}
E_6+D_7+A_5
\end{gathered}
$
& $38$
\\[8 pt]\hline

13 &
$
\begin{gathered}
E_6+A_{11}
\end{gathered}
$
& $38$
\\[8 pt]\hline

14 &
$
\begin{gathered}
3D_6
\end{gathered}
$
& $19$
\\[8 pt]\hline

15 &
$
\begin{gathered}
D_6+A_9+A_3
\end{gathered}
$
& $49$
\\[8 pt]\hline

16 &
$
A_{15}+A_3
$
& $24$
\\[8pt]\hline

17 &
$
2A_9
$
& $32$
\\[8pt]\hline

18 &
$
A_{17}+A_2
$
& $9$
\\[8pt]\hline

19 &
$
\begin{gathered}
3E_6+A_1
\end{gathered}
$
& $9$
\\[8 pt]\hline

20 &
$
D_7+A_{11}+A_1
$
& $24$
\\[8pt]\hline

21 &
$
A_{17}+A_1
$
& $18$
\\[8pt]\hline

22 &
$
\begin{gathered}
2D_5+A_7+A_1
\end{gathered}
$
& $67$
\\[8 pt]\hline

23 &
$
\begin{gathered}
2A_8+A_2
\end{gathered}
$
& $29$
\\[8 pt]\hline

24 &
$
\begin{gathered}
A_{12}+A_6
\end{gathered}
$
& $27$
\\[8 pt]\hline

25 &
$
A_{18}
$
& $9$
\\[8pt]\hline

26 &
$
\begin{gathered}
3A_6
\end{gathered}
$
& $23$
\\[8 pt]\hline

27 &
$
\begin{gathered}
D_4+3A_5
\end{gathered}
$
& $69$
\\[8 pt]\hline

&
\textbf{Total enhancements}
&
$\mathbf{699}$
\\[8pt]\hline

\caption{Rank 19 cases and enhancements for $\Lambda_R=A_2+A_1$.}
\label{tab: rank 3 A2+A1}\\

\end{longtable}
\endgroup
\begingroup
\renewcommand{\arraystretch}{1.6}
\setlength{\tabcolsep}{8pt}
\setlength\LTleft{-1 cm}
\setlength\LTright{\fill}

\begin{longtable}{
|>{\centering\arraybackslash}p{0.05\textwidth}|
>{\centering\arraybackslash}p{0.78\textwidth}|
>{\centering\arraybackslash}p{0.14\textwidth}|}

\hline
\# & Root system & Enhancements \\
\hline
\endfirsthead

\hline
\# & Root system & Enhancements \\
\hline
\endhead

\hline
\endhead

\endfoot

1 &
$2E_{8}+3A_{1}$
&11 \\[4pt] \hline

2 &
$\begin{gathered}
2E_{6}+A_{5}
\end{gathered}$
&71 \\[8 pt] \hline

3 &
$\begin{gathered}
D_{10}+D_{6}+3A_{1}
\end{gathered}$
&33 \\[8 pt] \hline

4 &
$\begin{gathered}
3D_{6}+A_{1}
\end{gathered}$
&71 \\[8 pt] \hline

5 &
$\begin{gathered}
E_{7}+D_{8}+4A_{1}
\end{gathered}$
&37 \\[8 pt] \hline

6 &
$\begin{gathered}
D_{8}+D_{6}+D_{4}+A_{1}
\end{gathered}$
&67 \\[8 pt] \hline

7 &
$D_{5}+A_{13}$
&31 \\[8pt] \hline

8 &
$A_{15}+3A_{1}$
&21 \\[8pt] \hline

9 &
$2E_{7}+D_{4}+A_{1}$
& 31\\[8pt] \hline

10 &
$D_{16}+3A_{1}$
&15 \\[8pt] \hline

11 &
$A_{15}+A_{3}+A_{1}$
&29 \\[8pt] \hline

12 &
$D_{12}+D_{6}+A_{1}$
&33 \\[8pt] \hline

13 &
$E_{7}+2D_{6}$
& 41\\[8pt] \hline

14 &
$E_{8}+D_{10}+A_{1}$
&23 \\[4pt] \hline

15 &
$E_{7}+D_{10}+2A_{1}$
& 29\\[8pt] \hline

16 &
$D_{10}+D_{8}+A_{1}$
& 31\\[8pt] \hline

17 &
$D_{18}+A_{1}$
&11 \\[4pt] \hline

18 &
$E_{7}+D_{12}$
& 17\\[4pt] \hline

19 &
$D_{14}+D_{4}+A_{1}$
&29 \\[8pt] \hline

20 &
$E_{8}+E_{7}+D_{4}$
&29 \\[4pt] \hline

21 &
$D_{5}+A_{7}+A_{5}$
&103 \\[8 pt] \hline

22 &
$A_{11}+A_{5}+A_{3}$
&43 \\[8pt] \hline

23 &
$E_{6}+A_{9}+A_{3}$
&53 \\[8pt] \hline

24 &
$A_{9}+A_{7}+2A_{1}$
&54 \\[8pt] \hline

25 &
$E_{6}+A_{11}+A_{1}$
&41 \\[8pt] \hline

26 &
$2A_{7}+A_{3}+A_{1}$
&93 \\[8 pt] \hline

27 &
$\begin{gathered}
2D_{6}+D_{4}+3A_{1}
\end{gathered}$
&99 \\[8 pt] \hline

28 &
$A_{17}+2A_{1}$
&14 \\[8pt] \hline

29 &
$D_{5}+A_{11}+A_{1}$
&71 \\[8pt] \hline

30 &
$2A_{9}+A_{1}$
&35 \\[8pt] \hline

31 &
$\begin{gathered}
3A_{5}+A_{3}
\end{gathered}$
&46 \\[8 pt] \hline

32 &
$\begin{gathered}
2D_{8}+3A_{1}
\end{gathered}$
&33 \\[8 pt] \hline

33 &
$\begin{gathered}
4D_{4}+3A_{1}
\end{gathered}$
&37 \\[8 pt] \hline

34 &
$D_{7}+A_{9}
$
&79 \\[8pt] \hline
&
\textbf{Total enhancements}
&\textbf{1461}
\\[8pt] \hline
\caption{\mbox{Root systems and number of enhancements for $\Lambda_R=3A_1$}}
\label{tab: rank 3 3A1}\\

\end{longtable}
\endgroup

\subsection{Rank 4}
The lists of root systems for $\Lambda_R=A_4,\;D_4$, and $2A_2$ are given in Tables \ref{tab: rank 4 A4}, \ref{tab: rank 4 2A2}, and \ref{tab: rank 4 D4} respectively.
\begingroup
\renewcommand{\arraystretch}{1.6}
\setlength{\tabcolsep}{8pt}
\setlength{\LTcapwidth}{\textwidth}
\setlength\LTleft{-1 cm}
\setlength\LTright{\fill}

\begin{longtable}{
|>{\centering\arraybackslash}p{0.05\textwidth}|
>{\centering\arraybackslash}p{0.78\textwidth}|
>{\centering\arraybackslash}p{0.14\textwidth}|}

\hline
\# & Root system & Enhancements \\
\hline
\endfirsthead

\hline
\# & Root system & Enhancements \\
\hline
\endhead

\endfoot

1 &
$2E_8+A_4$
& $9$
\\[8pt]\hline

2 &
$
E_7+D_{10}+A_2
$
& $29$
\\[8pt]\hline

3 &
$
2E_7+D_5
$
& $29$
\\[8pt]\hline

4 &
$
D_{16}+A_4+
$
& $12$
\\[8pt]\hline

5 &
$
\begin{gathered}
2D_8+A_3
\end{gathered}
$
& $28$
\\[8 pt]\hline

6 &
$
\begin{gathered}
3E_6+A_1
\end{gathered}
$
& $16$
\\[8 pt]\hline

7 &
$
\begin{gathered}
E_6+A_{11}
\end{gathered}
$
& $44$
\\[8 pt]\hline

8 &
$
\begin{gathered}
A_{17}+A_2
\end{gathered}
$
& $18$
\\[8 pt]\hline

9 &
$
\begin{gathered}
D_7+A_{11}+A_1
\end{gathered}
$
& $37$
\\[8 pt]\hline

10 &
$
\begin{gathered}
2D_5+A_7+A_2
\end{gathered}
$
& $83$
\\[8 pt]\hline

11 &
$
\begin{gathered}
E_6+D_7+A_6
\end{gathered}
$
& $41$
\\[8 pt]\hline

12 &
$
\begin{gathered}
D_6+A_9+A_4
\end{gathered}
$
& $56$
\\[8 pt]\hline

13 &
$
D_9+A_{10}
$
& $23$
\\[8pt]\hline

14 &
$
\begin{gathered}
D_{12}+D_7
\end{gathered}
$
& $20$
\\[8 pt]\hline

15 &
$
\begin{gathered}
D_4+A_{15}
\end{gathered}
$
& $28$
\\[8 pt]\hline

16 &
$
\begin{gathered}
E_7+A_{12}
\end{gathered}
$
& $20$
\\[8 pt]\hline

17 &
$
\begin{gathered}
D_5+2A_7
\end{gathered}
$
& $63$
\\[8 pt]\hline

18 &
$
E_8+D_{11}
$
& $13$
\\[8pt]\hline

19 &
$
\begin{gathered}
2A_8+A_3
\end{gathered}
$
& $38$
\\[8 pt]\hline

20 &
$
A_{12}+A_7
$
& $31$
\\[8pt]\hline

21 &
$
A_{19}
$
& $10$
\\[8pt]\hline

22 &
$
D_{19}
$
& $6$
\\[8pt]\hline

23 &
$
\begin{gathered}
3A_6+A_1
\end{gathered}
$
& $35$
\\[8 pt]\hline

24 &
$
\begin{gathered}
D_4+3A_5
\end{gathered}
$
& $123$
\\[8 pt]\hline

25 &
$
\begin{gathered}
5A_4
\end{gathered}
$
& $13$
\\[8 pt]\hline

26 &
$
\begin{gathered}
2A_9
\end{gathered}
$
& $59$
\\[8 pt]\hline

27 &
$
\begin{gathered}
3D_6
\end{gathered}
$
& $31$
\\[8 pt]\hline

&
\textbf{Total enhancements}
&
$\mathbf{915}$
\\[8pt]\hline

\caption{Root systems and number of enhancements for $\Lambda_R=A_4$}
\label{tab: rank 4 A4}\\

\end{longtable}
\endgroup
\begingroup
\renewcommand{\arraystretch}{1.6}
\setlength{\tabcolsep}{8pt}
\setlength\LTleft{-1 cm}
\setlength\LTright{\fill}
\begin{longtable}{|>{\centering\arraybackslash}p{0.05\textwidth}|
                  >{\centering\arraybackslash}p{0.78\textwidth}|
                  >{\centering\arraybackslash}p{0.14\textwidth}|}
\hline
\# & Root system & Enhancements \\
\hline
\endfirsthead

\hline
\# & Root system & Enhancements \\
\hline
\endhead

\endfoot


1 &
$
2E_{8}+2A_{2}
$
&
$9$
\\[8pt] \hline


2 &
$
E_{8}+2E_{6}
$
&
$9$
\\[8pt] \hline


3 &
$
\begin{gathered}E_{8}+D_{10}\end{gathered}
$
&
$25$
\\[8pt] \hline


4 &
$
\begin{gathered}2E_{7}+D_{4}\end{gathered}
$
&
$53$
\\[8pt] \hline


5 &
$
\begin{gathered}E_{7}+D_{10}+A_{2}\end{gathered}
$
&
$29$
\\[8pt] \hline


6 &
$
\begin{gathered}E_{7}+D_{7}+A_{5}\end{gathered}
$
&
$38$
\\[8pt] \hline


7 &
$
\begin{gathered}E_{7}+A_{11}\end{gathered}
$
&
$38$
\\[8pt] \hline


8 &
$
\begin{gathered}3E_{6}+A_{2}\end{gathered}
$
&
$9$
\\[8pt] \hline


9 &
$
\begin{gathered}2E_{6}+4A_{2}\end{gathered}
$
&
$40$
\\[8pt] \hline


10 &
$
\begin{gathered}E_{6}+D_{13}\end{gathered}
$
&
$15$
\\[8pt] \hline


11 &
$
\begin{gathered}E_{6}+D_{7}+A_{5}\end{gathered}
$
&
$73$
\\[8pt] \hline


12 &
$
\begin{gathered}E_{6}+D_{4}+A_{8}\end{gathered}
$
&
$83$
\\[8pt] \hline


13 &
$
\begin{gathered}E_{6}+A_{11}\end{gathered}
$
&
$78$
\\[8pt] \hline


14 &
$
\begin{gathered}D_{18}\end{gathered}
$
&
$12$
\\[8pt] \hline


15 &
$
\begin{gathered}D_{16}+2A_{2}\end{gathered}
$
&
$12$
\\[8pt] \hline


16 &
$
\begin{gathered}D_{12}+D_{6}\end{gathered}
$
&
$36$
\\[8pt] \hline


17 &
$
\begin{gathered}D_{10}+2A_{5}\end{gathered}
$
&
$30$
\\[8pt] \hline


18 &
$
\begin{gathered}2D_{9}\end{gathered}
$
&
$33$
\\[8pt] \hline


19 &
$
\begin{gathered}D_{9}+A_{9}\end{gathered}
$
&
$44$
\\[8pt] \hline


20 &
$
\begin{gathered}2D_{8}+2A_{1}\end{gathered}
$
&
$47$
\\[8pt] \hline


21 &
$
\begin{gathered}D_{8}+2D_{5}\end{gathered}
$
&
$74$
\\[8pt] \hline


22 &
$
\begin{gathered}D_{7}+A_{11}+A_{2}\end{gathered}
$
&
$24$
\\[8pt] \hline


23 &
$
\begin{gathered}D_{7}+A_{8}+2A_{2}\end{gathered}
$
&
$52$
\\[8pt] \hline


24 &
$
\begin{gathered}3D_{6}\end{gathered}
$
&
$80$
\\[8pt] \hline


25 &
$
\begin{gathered}2D_{6}+2A_{3}\end{gathered}
$
&
$142$
\\[8pt] \hline


26 &
$
\begin{gathered}D_{6}+A_{12}\end{gathered}
$
&
$45$
\\[8pt] \hline


27 &
$
\begin{gathered}D_{6}+A_{9}+A_{3}\end{gathered}
$
&
$94$
\\[8pt] \hline


28 &
$
\begin{gathered}D_{6}+2A_{6}\end{gathered}
$
&
$103$
\\[8pt] \hline


29 &
$
\begin{gathered}2D_{5}+A_{7}+A_{1}\end{gathered}
$
&
$126$
\\[8pt] \hline


30 &
$
\begin{gathered}2D_{5}+2A_{4}\end{gathered}
$
&
$163$
\\[8pt] \hline


31 &
$
\begin{gathered}D_{5}+A_{7}+A_{4}+2A_{1}\end{gathered}
$
&
$184$
\\[8pt] \hline


32 &
$
\begin{gathered}4D_{4}\end{gathered}
$
&
$181$
\\[8pt] \hline


33 &
$
\begin{gathered}D_{4}+A_{11}+2A_{2}\end{gathered}
$
&
$57$
\\[8pt] \hline


34 &
$
\begin{gathered}D_{4}+3A_{5}\end{gathered}
$
&
$123$
\\[8pt] \hline


35 &
$
\begin{gathered}D_{4}+2A_{5}+2A_{2}\end{gathered}
$
&
$273$
\\[8pt] \hline


36 &
$
\begin{gathered}A_{18}\end{gathered}
$
&
$19$
\\[8pt] \hline


37 &
$
\begin{gathered}A_{17}+A_{2}\end{gathered}
$
&
$18$
\\[8pt] \hline


38 &
$
\begin{gathered}A_{15}+A_{3}\end{gathered}
$
&
$48$
\\[8pt] \hline


39 &
$
\begin{gathered}A_{14}+A_{5}\end{gathered}
$
&
$27$
\\[8pt] \hline


40 &
$
\begin{gathered}A_{12}+A_{6}\end{gathered}
$
&
$55$
\\[8pt] \hline


41 &
$
\begin{gathered}2A_{9}\end{gathered}
$
&
$60$
\\[8pt] \hline


42 &
$
\begin{gathered}A_{9}+A_{6}+A_{3}\end{gathered}
$
&
$109$
\\[8pt] \hline


43 &
$
\begin{gathered}2A_{8}+A_{2}\end{gathered}
$
&
$59$
\\[8pt] \hline


44 &
$
\begin{gathered}A_{8}+2A_{5}\end{gathered}
$
&
$122$
\\[8pt] \hline


45 &
$
\begin{gathered}2A_{7}+4A_{1}\end{gathered}
$
&
$124$
\\[8pt] \hline


46 &
$
\begin{gathered}3A_{6}\end{gathered}
$
&
$127$
\\[8pt] \hline


47 &
$
\begin{gathered}2A_{6}+2A_{3}\end{gathered}
$
&
$199$
\\[8pt] \hline


48 &
$
\begin{gathered}3A_{5}+A_{2}\end{gathered}
$
&
$342$
\\[8pt] \hline


49 &
$
\begin{gathered}4A_{4}+2A_{1}\end{gathered}
$
&
$364$
\\[8pt] \hline


50 &
$
\begin{gathered}6A_{3}\end{gathered}
$
&
$173$
\\[8pt] \hline


51 &
$
\begin{gathered}10A_{2}\end{gathered}
$
&
$73$
\\[8pt] \hline


&
\textbf{Total enhancements}
&
$\mathbf{4353}$
\\[8pt] \hline

\caption{\mbox{Root systems and number of enhancements for $\Lambda_R=2A_2$.}}
\label{tab: rank 4 2A2}\\
\end{longtable}
\endgroup

\begingroup
\renewcommand{\arraystretch}{1.6}
\setlength{\tabcolsep}{8pt}
\setlength\LTleft{-1 cm}
\setlength\LTright{\fill}

\begin{longtable}{
|>{\centering\arraybackslash}p{0.05\textwidth}|
>{\centering\arraybackslash}p{0.78\textwidth}|
>{\centering\arraybackslash}p{0.14\textwidth}|}

\hline
\# & Root system & Enhancements \\
\hline
\endfirsthead

\hline
\# & Root system & Enhancements \\
\hline
\endhead

\endfoot


1 &
$2E_8+D_4$
&17 \\[4pt] \hline


2 &
$E_8+D_{12}$
&14 \\[4pt] \hline


3 &
$
2E_7+D_6
$
& 20\\[8pt] \hline


4 &
$
\begin{gathered}
E_7+D_{10}
\end{gathered}
$
&24 \\[8 pt] \hline


5 &
$
\begin{gathered}
3E_6
\end{gathered}
$
&23 \\[8 pt] \hline


6 &
$
E_6+A_{11}+A_3
$
&33 \\[8pt] \hline


7 &
$D_{20}$
&7 \\[4pt] \hline


8 &
$
D_{16}+D_4
$
&22 \\[8pt] \hline


9 &
$
D_{12}+D_8
$
&22 \\[8pt] \hline


10 &
$
\begin{gathered}
2D_8+D_4
\end{gathered}
$
&33 \\[8 pt] \hline


11 &
$
D_7+A_{11}
$
&51 \\[8pt] \hline


12 &
$
\begin{gathered}
3D_6+2A_1
\end{gathered}
$
&69 \\[8 pt] \hline


13 &
$
D_5+A_{15}
$
&21 \\[8pt] \hline


14 &
$
\begin{gathered}
D_5+2A_7
\end{gathered}
$
&69 \\[8 pt] \hline


15 &
$
A_{17}+3A_1
$
&11 \\[8pt] \hline


16 &
$
2A_9+2A_1
$
&32 \\[8pt] \hline


17 &
$
\begin{gathered}
5D_4
\end{gathered}
$
&111 \\[8 pt] \hline


18 &
$
\begin{gathered}
4A_5
\end{gathered}
$
&31 \\[8 pt] \hline

&
\textbf{Total enhancements}
&\textbf{610}
\\[8pt] \hline

\caption{Root systems and number of enhancements for $\Lambda_R=D_4$.}
\label{tab: rank 4 D4}\\

\end{longtable}
\endgroup
\subsection{Hessians}
The following three tables tabulate the non-knife edge points and the value of the potential $V$, the smallest eigenvalue of the Hessian (with respect to the canonical $\s_{ia}$ moduli) that we call $\la_{\text{min}}$, and the ratio $\la_{\text{min}}/V$. Those points whose enhancements, and the values of $V$ and $\la_{\text{min}}$ are the same, are placed in the same class. We do this to find the lower bound on the number of points which are manifestly physically inequivalent. This lower bound is the number of classes, which turns out to be $2,\;11,$ and $38$ for $T^2,\;T^3,$ and $T^4$ respectively. We don't calculate the Hessians for the non-knife edge extrema for $S^1$ as they are already provided in \cite{Fraiman:2023cpa}.
\begingroup
\renewcommand{\arraystretch}{1.6}
\setlength{\tabcolsep}{8pt}
\setlength\LTleft{-1 cm}
\setlength\LTright{\fill}
\begin{longtable}{|>{\centering\arraybackslash}p{0.15\textwidth}|>{\centering\arraybackslash}p{0.08\textwidth}|>{\centering\arraybackslash}p{0.2\textwidth}|>{\centering\arraybackslash}p{0.12\textwidth}|>{\centering\arraybackslash}p{0.12\textwidth}|>{\centering\arraybackslash}p{0.11\textwidth}|>{\centering\arraybackslash}p{0.045\textwidth}|}
\hline
Root system & $\Lambda_R$ & Enhancement & $10^6\lambda_{\min}$ & $10^6V$ & $\lambda_{\min}/V$ & Class \\
\hline
\endfirsthead

\hline
Root system & $\Lambda_R$ & Enhancement & $10^6\lambda_{\min}$ & $10^6V$ & $\lambda_{\min}/V$ & Class \\
\hline
\endhead

\endfoot

$2E_{8}+A_{2}$ & $A_{2}$ & $2D_{8}+A_{2}$ & $-154.367$ & $196.863488$ & $-0.784$ & 1 \\[8pt] \hline

$D_{16}+A_{2}$ & $A_{2}$ & $2D_{8}+A_{2}$ & $-154.367$ & $196.863488$ & $-0.784$ & 1 \\[8pt] \hline

$2E_{8}+2A_{1}$ & $2A_{1}$ & $2D_{8}+2A_{1}$ & $-155.584$ & $211.623906$ & $-0.735$ & 2 \\[8pt] \hline

$D_{16}+2A_{1}$ & $2A_{1}$ & $2D_{8}+2A_{1}$ & $-155.584$ & $211.623906$ & $-0.735$ & 2 \\[8pt] \hline

\caption{\mbox{Hessians for the non-knife-edge points for $T^2$, sorted by class.} These results match with the last two rows in Table 1 of \cite{honda2026}. }\label{tab:nonknife-T2}\\
\end{longtable}
\endgroup
\begingroup
\renewcommand{\arraystretch}{1.6}
\setlength{\tabcolsep}{8pt}
\setlength\LTleft{-1 cm}
\setlength\LTright{\fill}
\begin{longtable}{|>{\centering\arraybackslash}p{0.15\textwidth}|>{\centering\arraybackslash}p{0.08\textwidth}|>{\centering\arraybackslash}p{0.2\textwidth}|>{\centering\arraybackslash}p{0.12\textwidth}|>{\centering\arraybackslash}p{0.12\textwidth}|>{\centering\arraybackslash}p{0.11\textwidth}|>{\centering\arraybackslash}p{0.045\textwidth}|}
\hline
Root system & $\Lambda_R$ & Enhancement & $10^4\lambda_{\min}$ & $10^4V$ & $\lambda_{\min}/V$ & Class \\
\hline
\endfirsthead

\hline
Root system & $\Lambda_R$ & Enhancement & $10^4\lambda_{\min}$ & $10^4V$ & $\lambda_{\min}/V$ & Class \\
\hline
\endhead

\endfoot

$2E_{8}+A_{3}$ & $A_{3}$ & $2D_{8}+A_{3}$ & $-11.922$ & $13.459$ & $-0.886$ & 1 \\[8pt] \hline

$D_{16}+A_{3}$ & $A_{3}$ & $2D_{8}+A_{3}$ & $-11.922$ & $13.459$ & $-0.886$ & 1 \\[8pt] \hline

$2E_{8}+A_{2}+A_{1}$ & $A_{2}+A_{1}$ & $2D_{8}+A_{2}+A_{1}$ & $-12.243$ & $14.805$ & $-0.827$ & 2 \\[8pt] \hline

$D_{16}+A_{2}+A_{1}$ & $A_{2}+A_{1}$ & $2D_{8}+A_{2}+A_{1}$ & $-12.243$ & $14.805$ & $-0.827$ & 2 \\[8pt] \hline

$2E_{8}+A_{2}+A_{1}$ & $A_{2}+A_{1}$ & \textcolor{red}{$2D_{8}+2A_{1}$} & $-20.833$ & $0.598$ & $-34.864$ & 3 \\[8pt] \hline

$D_{16}+A_{2}+A_{1}$ & $A_{2}+A_{1}$ & \textcolor{red}{$2D_{8}+2A_{1}$} & $-20.833$ & $0.598$ & $-34.864$ & 3 \\[8pt] \hline

$2D_{8}+2A_{1}$ & $A_{2}+A_{1}$ & \textcolor{red}{$2D_{8}+2A_{1}$} & $-20.833$ & $0.598$ & $-34.864$ & 3 \\[8pt] \hline

$2D_{8}+2A_{1}$ & $A_{2}+A_{1}$ & \textcolor{red}{$2D_{8}+2A_{1}$} & $-20.833$ & $0.598$ & $-34.864$ & 3 \\[8pt] \hline

$E_{8}+D_{10}$ & $A_{2}+A_{1}$ & \textcolor{red}{$D_{10}+D_{8}$} & $-32.884$ & $7.701$ & $-4.270$ & 4 \\[8pt] \hline

$D_{18}$ & $A_{2}+A_{1}$ & \textcolor{red}{$D_{10}+D_{8}$} & $-32.884$ & $7.701$ & $-4.270$ & 4 \\[8pt] \hline

$E_{8}+D_{10}$ & $A_{2}+A_{1}$ & \textcolor{red}{$D_{8}+D_{6}+D_{4}$} & $-32.884$ & $7.701$ & $-4.270$ & 5 \\[8pt] \hline

$D_{12}+D_{6}$ & $A_{2}+A_{1}$ & \textcolor{red}{$D_{8}+D_{6}+D_{4}$} & $-32.884$ & $7.701$ & $-4.270$ & 5 \\[8pt] \hline

$2E_{7}+D_{4}$ & $A_{2}+A_{1}$ & \textcolor{red}{$2D_{6}+D_{4}+2A_{1}$} & $-32.884$ & $7.701$ & $-4.270$ & 6 \\[8pt] \hline

$D_{12}+D_{6}$ & $A_{2}+A_{1}$ & \textcolor{red}{$2D_{6}+D_{4}+2A_{1}$} & $-32.884$ & $7.701$ & $-4.270$ & 6 \\[8pt] \hline

$2E_{7}+D_{4}$ & $A_{2}+A_{1}$ & \textcolor{red}{$2A_{7}+A_{3}$} & $-32.884$ & $7.701$ & $-4.270$ & 7 \\[8pt] \hline

$A_{15}+A_{3}$ & $A_{2}+A_{1}$ & \textcolor{red}{$2A_{7}+A_{3}$} & $-32.884$ & $7.701$ & $-4.270$ & 7 \\[8pt] \hline

$E_{7}+D_{10}+A_{1}$ & $A_{2}+A_{1}$ & \textcolor{red}{$2D_{5}+A_{7}+A_{1}$} & $-7.270$ & $0.598$ & $-12.166$ & 8 \\[8pt] \hline

$2D_{5}+A_{7}+A_{1}$ & $A_{2}+A_{1}$ & \textcolor{red}{$2D_{5}+A_{7}+A_{1}$} & $-7.270$ & $0.598$ & $-12.166$ & 8 \\[8pt] \hline

$2D_{8}+2A_{1}$ & $A_{2}+A_{1}$ & \textcolor{red}{$4D_{4}+2A_{1}$} & $-12.992$ & $7.701$ & $-1.687$ & 9 \\[8pt] \hline

$2E_{8}+3A_{1}$ & $3A_{1}$ & $2D_{8}+3A_{1}$ & $-12.195$ & $15.928$ & $-0.766$ & 10 \\[8pt] \hline

$D_{16}+3A_{1}$ & $3A_{1}$ & $2D_{8}+3A_{1}$ & $-12.195$ & $15.928$ & $-0.766$ & 10 \\[8pt] \hline

$D_{14}+D_{4}+A_{1}$ & $3A_{1}$ & $D_{8}+D_{6}+D_{4}+A_{1}$ & $-11.285$ & $9.942$ & $-1.135$ & 11 \\[8pt] \hline

$E_{8}+E_{7}+D_{4}$ & $3A_{1}$ & $D_{8}+D_{6}+D_{4}+A_{1}$ & $-11.285$ & $9.942$ & $-1.135$ & 11 \\[8pt] \hline

\caption{Hessians for the non-knife-edge points for $T^3$, sorted by class. Red enhancements are the non-maximal enhancements.}\label{tab:nonknife-T3}\\
\end{longtable}
\endgroup

\begingroup
\renewcommand{\arraystretch}{1.6}
\setlength{\tabcolsep}{8pt}
\setlength\LTleft{-1 cm}
\setlength\LTright{\fill}
\begin{longtable}{|>{\centering\arraybackslash}p{0.15\textwidth}|>{\centering\arraybackslash}p{0.08\textwidth}|>{\centering\arraybackslash}p{0.2\textwidth}|>{\centering\arraybackslash}p{0.12\textwidth}|>{\centering\arraybackslash}p{0.12\textwidth}|>{\centering\arraybackslash}p{0.11\textwidth}|>{\centering\arraybackslash}p{0.045\textwidth}|}
\hline
Root system & $\Lambda_R$ & Enhancement & $10^4\lambda_{\min}$ & $10^4V$ & $\lambda_{\min}/V$ & Class \\
\hline
\endfirsthead

\hline
Root system & $\Lambda_R$ & Enhancement & $10^4\lambda_{\min}$ & $10^4V$ & $\lambda_{\min}/V$ & Class \\
\hline
\endhead

\endfoot

$2E_{8}+A_{4}$ & $A_{4}$ & $2D_{8}+A_{4}$ & $-91.164$ & $92.982$ & $-0.981$ & 1 \\[8pt] \hline

$2E_{8}+A_{4}$ & $A_{4}$ & \textcolor{red}{$2D_{8}+A_{3}$} & $-110.410$ & $-6.872$ & $16.066$ & 2 \\[8pt] \hline

$D_{16}+A_{4}$ & $A_{4}$ & \textcolor{red}{$2D_{8}+A_{3}$} & $-110.410$ & $-6.872$ & $16.066$ & 2 \\[8pt] \hline

$2D_{8}+A_{3}$ & $A_{4}$ & \textcolor{red}{$2D_{8}+A_{3}$} & $-110.410$ & $-6.872$ & $16.066$ & 2 \\[8pt] \hline

$2D_{8}+A_{3}$ & $A_{4}$ & \textcolor{red}{$2D_{8}+A_{3}$} & $-110.410$ & $-6.872$ & $16.066$ & 2 \\[8pt] \hline

$E_{7}+D_{10}+A_{2}$ & $A_{4}$ & \textcolor{red}{$2D_{5}+A_{7}+A_{2}$} & $-37.306$ & $-6.872$ & $5.429$ & 3 \\[8pt] \hline

$2D_{5}+A_{7}+A_{2}$ & $A_{4}$ & \textcolor{red}{$2D_{5}+A_{7}+A_{2}$} & $-37.306$ & $-6.872$ & $5.429$ & 3 \\[8pt] \hline

$2E_{7}+D_{5}$ & $A_{4}$ & \textcolor{red}{$2D_{6}+D_{5}+2A_{1}$} & $-200.332$ & $43.055$ & $-4.653$ & 4 \\[8pt] \hline

$D_{12}+D_{7}$ & $A_{4}$ & \textcolor{red}{$2D_{6}+D_{5}+2A_{1}$} & $-200.332$ & $43.055$ & $-4.653$ & 4 \\[8pt] \hline

$2E_{7}+D_{5}$ & $A_{4}$ & \textcolor{red}{$D_{4}+2A_{7}$} & $-200.332$ & $43.055$ & $-4.653$ & 5 \\[8pt] \hline

$D_{4}+A_{15}$ & $A_{4}$ & \textcolor{red}{$D_{4}+2A_{7}$} & $-200.332$ & $43.055$ & $-4.653$ & 5 \\[8pt] \hline

$D_{16}+A_{4}$ & $A_{4}$ & $2D_{8}+A_{4}$ & $-91.164$ & $92.982$ & $-0.981$ & 6 \\[8pt] \hline

$2D_{8}+A_{3}$ & $A_{4}$ & \textcolor{red}{$4D_{4}+A_{3}$} & $-64.235$ & $43.055$ & $-1.492$ & 7 \\[8pt] \hline

$D_{12}+D_{7}$ & $A_{4}$ & \textcolor{red}{$D_{8}+D_{7}+D_{4}$} & $-200.332$ & $43.055$ & $-4.653$ & 8 \\[8pt] \hline

$E_{8}+D_{11}$ & $A_{4}$ & \textcolor{red}{$D_{8}+D_{7}+D_{4}$} & $-200.332$ & $43.055$ & $-4.653$ & 8 \\[8pt] \hline

$E_{8}+D_{11}$ & $A_{4}$ & \textcolor{red}{$D_{11}+D_{8}$} & $-200.332$ & $43.055$ & $-4.653$ & 9 \\[8pt] \hline

$D_{19}$ & $A_{4}$ & \textcolor{red}{$D_{11}+D_{8}$} & $-200.332$ & $43.055$ & $-4.653$ & 9 \\[8pt] \hline

$2E_{8}+2A_{2}$ & $2A_{2}$ & $2D_{8}+2A_{2}$ & $-91.541$ & $105.354$ & $-0.869$ & 10 \\[8pt] \hline

$D_{16}+2A_{2}$ & $2A_{2}$ & $2D_{8}+2A_{2}$ & $-91.541$ & $105.354$ & $-0.869$ & 10 \\[8pt] \hline

$2E_{8}+2A_{2}$ & $2A_{2}$ & \textcolor{red}{$2D_{8}+2A_{1}$} & $-73.729$ & $8.331$ & $-8.850$ & 11 \\[8pt] \hline

$D_{16}+2A_{2}$ & $2A_{2}$ & \textcolor{red}{$2D_{8}+2A_{1}$} & $-73.729$ & $8.331$ & $-8.850$ & 11 \\[8pt] \hline

$2D_{8}+2A_{1}$ & $2A_{2}$ & \textcolor{red}{$2D_{8}+2A_{1}$} & $-73.729$ & $8.331$ & $-8.850$ & 11 \\[8pt] \hline

$2D_{8}+2A_{1}$ & $2A_{2}$ & \textcolor{red}{$2D_{8}+2A_{1}$} & $-73.729$ & $8.331$ & $-8.850$ & 11 \\[8pt] \hline

$E_{8}+2E_{6}$ & $2A_{2}$ & \textcolor{red}{$E_{6}+D_{8}+D_{5}$} & $-94.064$ & $32.587$ & $-2.887$ & 12 \\[8pt] \hline

$E_{6}+D_{13}$ & $2A_{2}$ & \textcolor{red}{$E_{6}+D_{8}+D_{5}$} & $-94.064$ & $32.587$ & $-2.887$ & 12 \\[8pt] \hline

$E_{8}+2E_{6}$ & $2A_{2}$ & $D_{8}+2A_{5}+2A_{1}$ & $-199.864$ & $32.587$ & $-6.134$ & 13 \\[8pt] \hline

$D_{10}+2A_{5}$ & $2A_{2}$ & $D_{8}+2A_{5}+2A_{1}$ & $-199.864$ & $32.587$ & $-6.134$ & 13 \\[8pt] \hline

$E_{8}+D_{10}$ & $2A_{2}$ & \textcolor{red}{$D_{10}+D_{8}$} & $-163.756$ & $56.842$ & $-2.881$ & 14 \\[8pt] \hline

$D_{18}$ & $2A_{2}$ & \textcolor{red}{$D_{10}+D_{8}$} & $-163.756$ & $56.842$ & $-2.881$ & 14 \\[8pt] \hline

$E_{8}+D_{10}$ & $2A_{2}$ & \textcolor{red}{$D_{8}+D_{6}+D_{4}$} & $-163.756$ & $56.842$ & $-2.881$ & 15 \\[8pt] \hline

$D_{12}+D_{6}$ & $2A_{2}$ & \textcolor{red}{$D_{8}+D_{6}+D_{4}$} & $-163.756$ & $56.842$ & $-2.881$ & 15 \\[8pt] \hline

$E_{8}+D_{10}$ & $2A_{2}$ & \textcolor{red}{$D_{8}+2D_{5}$} & $-24.372$ & $8.331$ & $-2.926$ & 16 \\[8pt] \hline

$D_{8}+2D_{5}$ & $2A_{2}$ & \textcolor{red}{$D_{8}+2D_{5}$} & $-24.372$ & $8.331$ & $-2.926$ & 16 \\[8pt] \hline

$2E_{7}+D_{4}$ & $2A_{2}$ & \textcolor{red}{$2D_{6}+D_{4}+2A_{1}$} & $-163.756$ & $56.842$ & $-2.881$ & 17 \\[8pt] \hline

$D_{12}+D_{6}$ & $2A_{2}$ & \textcolor{red}{$2D_{6}+D_{4}+2A_{1}$} & $-163.756$ & $56.842$ & $-2.881$ & 17 \\[8pt] \hline

$2E_{7}+D_{4}$ & $2A_{2}$ & \textcolor{red}{$2A_{7}+A_{3}$} & $-163.756$ & $56.842$ & $-2.881$ & 18 \\[8pt] \hline

$A_{15}+A_{3}$ & $2A_{2}$ & \textcolor{red}{$2A_{7}+A_{3}$} & $-163.756$ & $56.842$ & $-2.881$ & 18 \\[8pt] \hline

$2E_{7}+D_{4}$ & $2A_{2}$ & \textcolor{red}{$2A_{7}+4A_{1}$} & $-24.372$ & $8.331$ & $-2.926$ & 19 \\[8pt] \hline

$2A_{7}+4A_{1}$ & $2A_{2}$ & \textcolor{red}{$2A_{7}+4A_{1}$} & $-24.372$ & $8.331$ & $-2.926$ & 19 \\[8pt] \hline

$E_{7}+D_{10}+A_{2}$ & $2A_{2}$ & \textcolor{red}{$2D_{5}+A_{7}+A_{1}$} & $-73.729$ & $8.331$ & $-8.850$ & 20 \\[8pt] \hline

$2D_{5}+A_{7}+A_{1}$ & $2A_{2}$ & \textcolor{red}{$2D_{5}+A_{7}+A_{1}$} & $-73.729$ & $8.331$ & $-8.850$ & 20 \\[8pt] \hline

$E_{7}+D_{7}+A_{5}$ & $2A_{2}$ & \textcolor{red}{$D_{7}+D_{6}+A_{5}+A_{1}$} & $-199.864$ & $32.587$ & $-6.134$ & 21 \\[8pt] \hline

$E_{6}+D_{13}$ & $2A_{2}$ & \textcolor{red}{$D_{7}+D_{6}+A_{5}+A_{1}$} & $-199.864$ & $32.587$ & $-6.134$ & 21 \\[8pt] \hline

$E_{7}+D_{7}+A_{5}$ & $2A_{2}$ & \textcolor{red}{$D_{6}+D_{4}+A_{5}+A_{3}+A_{1}$} & $-146.964$ & $32.587$ & $-4.510$ & 22 \\[8pt] \hline

$D_{10}+2A_{5}$ & $2A_{2}$ & \textcolor{red}{$D_{6}+D_{4}+A_{5}+A_{3}+A_{1}$} & $-146.964$ & $32.587$ & $-4.510$ & 22 \\[8pt] \hline

$D_{10}+2A_{5}$ & $2A_{2}$ & \textcolor{red}{$D_{6}+D_{4}+A_{5}+A_{3}+A_{1}$} & $-146.964$ & $32.587$ & $-4.510$ & 22 \\[8pt] \hline

$E_{7}+D_{7}+A_{5}$ & $2A_{2}$ & \textcolor{red}{$D_{4}+A_{7}+A_{3}+2A_{2}$} & $-146.964$ & $32.587$ & $-4.510$ & 23 \\[8pt] \hline

$D_{4}+A_{11}+2A_{2}$ & $2A_{2}$ & \textcolor{red}{$D_{4}+A_{7}+A_{3}+2A_{2}$} & $-146.964$ & $32.587$ & $-4.510$ & 23 \\[8pt] \hline

$E_{7}+D_{7}+A_{5}$ & $2A_{2}$ & \textcolor{red}{$A_{7}+A_{6}+A_{5}$} & $-173.414$ & $32.587$ & $-5.322$ & 24 \\[8pt] \hline

$A_{14}+A_{5}$ & $2A_{2}$ & \textcolor{red}{$A_{7}+A_{6}+A_{5}$} & $-173.414$ & $32.587$ & $-5.322$ & 24 \\[8pt] \hline

$2E_{6}+4A_{2}$ & $2A_{2}$ & \textcolor{red}{$2D_{5}+4A_{2}$} & $-146.964$ & $32.587$ & $-4.510$ & 25 \\[8pt] \hline

$D_{10}+2A_{5}$ & $2A_{2}$ & \textcolor{red}{$2D_{5}+4A_{2}$} & $-146.964$ & $32.587$ & $-4.510$ & 25 \\[8pt] \hline

$2E_{6}+4A_{2}$ & $2A_{2}$ & \textcolor{red}{$2A_{5}+2A_{2}+4A_{1}$} & $-120.514$ & $32.587$ & $-3.699$ & 26 \\[8pt] \hline

$2E_{6}+4A_{2}$ & $2A_{2}$ & \textcolor{red}{$2A_{5}+2A_{2}+4A_{1}$} & $-120.514$ & $32.587$ & $-3.699$ & 26 \\[8pt] \hline

$D_{4}+A_{11}+2A_{2}$ & $2A_{2}$ & \textcolor{red}{$2A_{5}+2A_{2}+4A_{1}$} & $-120.514$ & $32.587$ & $-3.699$ & 26 \\[8pt] \hline

$D_{18}$ & $2A_{2}$ & \textcolor{red}{$2D_{9}$} & $-24.372$ & $8.331$ & $-2.926$ & 27 \\[8pt] \hline

$2D_{9}$ & $2A_{2}$ & \textcolor{red}{$2D_{9}$} & $-24.372$ & $8.331$ & $-2.926$ & 27 \\[8pt] \hline

$D_{12}+D_{6}$ & $2A_{2}$ & \textcolor{red}{$2D_{6}+2A_{3}$} & $-24.372$ & $8.331$ & $-2.926$ & 28 \\[8pt] \hline

$2D_{6}+2A_{3}$ & $2A_{2}$ & \textcolor{red}{$2D_{6}+2A_{3}$} & $-24.372$ & $8.331$ & $-2.926$ & 28 \\[8pt] \hline

$2D_{9}$ & $2A_{2}$ & \textcolor{red}{$2D_{5}+2D_{4}$} & $-94.064$ & $32.587$ & $-2.887$ & 29 \\[8pt] \hline

$D_{8}+2D_{5}$ & $2A_{2}$ & \textcolor{red}{$2D_{5}+2D_{4}$} & $-94.064$ & $32.587$ & $-2.887$ & 29 \\[8pt] \hline

$2D_{8}+2A_{1}$ & $2A_{2}$ & \textcolor{red}{$4D_{4}$} & $-24.372$ & $8.331$ & $-2.926$ & 30 \\[8pt] \hline

$4D_{4}$ & $2A_{2}$ & \textcolor{red}{$4D_{4}$} & $-24.372$ & $8.331$ & $-2.926$ & 30 \\[8pt] \hline

$4D_{4}$ & $2A_{2}$ & \textcolor{red}{$4D_{4}$} & $-24.372$ & $8.331$ & $-2.926$ & 30 \\[8pt] \hline

$4D_{4}$ & $2A_{2}$ & \textcolor{red}{$4D_{4}$} & $-24.372$ & $8.331$ & $-2.926$ & 30 \\[8pt] \hline

$4D_{4}$ & $2A_{2}$ & \textcolor{red}{$4D_{4}$} & $-24.372$ & $8.331$ & $-2.926$ & 30 \\[8pt] \hline

$4D_{4}$ & $2A_{2}$ & \textcolor{red}{$4D_{4}$} & $-24.372$ & $8.331$ & $-2.926$ & 30 \\[8pt] \hline

$4D_{4}$ & $2A_{2}$ & \textcolor{red}{$4D_{4}$} & $-24.372$ & $8.331$ & $-2.926$ & 30 \\[8pt] \hline

$4D_{4}$ & $2A_{2}$ & \textcolor{red}{$4D_{4}$} & $-24.372$ & $8.331$ & $-2.926$ & 30 \\[8pt] \hline

$4D_{4}$ & $2A_{2}$ & \textcolor{red}{$4D_{4}$} & $-24.372$ & $8.331$ & $-2.926$ & 30 \\[8pt] \hline

$4D_{4}$ & $2A_{2}$ & \textcolor{red}{$4D_{4}$} & $-24.372$ & $8.331$ & $-2.926$ & 30 \\[8pt] \hline

$2D_{8}+2A_{1}$ & $2A_{2}$ & \textcolor{red}{$4D_{4}+2A_{1}$} & $-57.956$ & $56.842$ & $-1.020$ & 31 \\[8pt] \hline

$D_{8}+2D_{5}$ & $2A_{2}$ & \textcolor{red}{$2D_{4}+2A_{3}+4A_{1}$} & $-94.064$ & $32.587$ & $-2.887$ & 32 \\[8pt] \hline

$2D_{6}+2A_{3}$ & $2A_{2}$ & \textcolor{red}{$2D_{4}+2A_{3}+4A_{1}$} & $-94.064$ & $32.587$ & $-2.887$ & 32 \\[8pt] \hline

$D_{7}+A_{8}+2A_{2}$ & $2A_{2}$ & \textcolor{red}{$D_{4}+A_{4}+2A_{3}+2A_{2}$} & $-120.514$ & $32.587$ & $-3.699$ & 33 \\[8pt] \hline

$3D_{6}$ & $2A_{2}$ & \textcolor{red}{$6A_{3}$} & $-24.372$ & $8.331$ & $-2.926$ & 34 \\[8pt] \hline

$6A_{3}$ & $2A_{2}$ & \textcolor{red}{$6A_{3}$} & $-24.372$ & $8.331$ & $-2.926$ & 34 \\[8pt] \hline

$2A_{7}+4A_{1}$ & $2A_{2}$ & \textcolor{red}{$4A_{3}+4A_{1}$} & $-94.064$ & $32.587$ & $-2.887$ & 35 \\[8pt] \hline

$2D_{6}+2A_{3}$ & $2A_{2}$ & \textcolor{red}{$4A_{3}+4A_{1}$} & $-94.064$ & $32.587$ & $-2.887$ & 35 \\[8pt] \hline

$3A_{6}$ & $2A_{2}$ & \textcolor{red}{$3A_{6}$} & $-19.322$ & $-17.265$ & $1.119$ & 36 \\[8pt] \hline

$3A_{6}$ & $2A_{2}$ & \textcolor{red}{$3A_{6}$} & $-19.322$ & $-17.265$ & $1.119$ & 36 \\[8pt] \hline

$4D_{4}$ & $2A_{2}$ & \textcolor{red}{$16A_{1}$} & $-41.164$ & $32.587$ & $-1.263$ & 37 \\[8pt] \hline

$2E_{8}+D_{4}$ & $D_{4}$ & $2D_{8}+D_{4}$ & $-91.936$ & $87.430$ & $-1.052$ & 38 \\[8pt] \hline

$D_{16}+D_{4}$ & $D_{4}$ & $2D_{8}+D_{4}$ & $-91.936$ & $87.430$ & $-1.052$ & 38 \\[8pt] \hline

\caption{Hessians for the non-knife-edge points for $T^4$, sorted by class. Red enhancements are the non-maximal enhancements.}\label{tab:nonknife-T4}\\
\end{longtable}
\endgroup
\printbibliography
\end{document}